\documentclass[%
 aip,
 pof,
 amsmath,
 amssymb,
 floatfix,%
]{revtex4-1}

\usepackage{
  amssymb, mathtools, bm, nicefrac, amsthm, 
  dsfont, 
  commath,
  color, tikz, pgfplots, xintexpr, booktabs, multirow,
  graphicx,
  dcolumn, mathptmx, etoolbox, enumitem
}
\usepackage{printlen}
\usepackage{standalone}
\usepackage[normalem]{ulem} 
\usepackage{soul} 

\pgfrealjobname{ms}

\usepackage{siunitx}
\usepackage[caption=false]{subfig}
\usepackage[hidelinks, colorlinks]{hyperref}

\usepgfplotslibrary{colormaps}
\usetikzlibrary{calc}
\usetikzlibrary{shapes.geometric}
\usetikzlibrary{positioning}

\pgfplotsset{%
  compat = newest,
  colormap={bled}{
			rgb={0.02,0.38129999999999997,0.99809999999999999}
			rgb={0.02000006,0.42426776799999999,0.96906968999999998}
			rgb={0.02,0.46723376300000002,0.94003304300000001}
			rgb={0.02,0.51019999999999999,0.91100000000000003}
			rgb={0.02000006,0.54640149400000004,0.87266943799999996}
			rgb={ 0.02,0.58260036199999998,0.83433294999999996}
			rgb={ 0.02,0.61880000000000002,0.79600000000000004}
			rgb={ 0.02000006,0.65253515600000001,0.74980243400000002}
			rgb={ 0.02,0.68626700399999996,0.70359953799999997}
			rgb={ 0.02,0.71999999999999997,0.65739999999999998}
			rgb={ 0.02000006,0.757035456,0.60373535899999997}
			rgb={ 0.02,0.79406703700000003,0.55006613000000004}
			rgb={  0.02,0.83109999999999995,0.49640000000000001}
			rgb={ 0.021354336738172372,0.86453685552616311,0.42855794607611591}
			rgb={ 0.023312914349117714,0.89799935992448399,0.36073871343115577}
			rgb={ 0.015976108242848862,0.9310479513349017,0.29256318150880922}
			rgb={ 0.27421074700988196,0.95256296099508297,0.15356836602739213}
			rgb={ 0.49335462816816988,0.96190386253094817,0.11119493614749336}
			rgb={ 0.64390000000000003,0.97729999999999995,0.046899999999999997}
			rgb={ 0.76240181299999998,0.98466959099999996,0.034600153000000002}
			rgb={ 0.88090118500000003,0.99203340699999998,0.022299876999999999}
			rgb={ 0.99952854326271467,0.99951937067814922,0.0134884641450013}
			rgb={ 0.99940299799999999,0.95503637600000002,0.079066628}
			rgb={0.99939999999999996,0.910666223,0.148134024}
			rgb={0.99939999999999996,0.86629999999999996,0.2172}
			rgb={0.99926966500000003,0.81803598099999997,0.21720065199999999}
			rgb={0.99913333199999999,0.76976618399999996,0.2172}
			rgb={0.999,0.72150000000000003,0.2172}
			rgb={0.99913633000000002,0.673435546,0.21720065199999999}
			rgb={0.99926666799999997,0.62536618600000005,0.2172}
			rgb={0.99939999999999996,0.57730000000000004,0.2172}
			rgb={0.99940299799999999,0.52106845499999999,0.21720065199999999}
			rgb={0.99939999999999996,0.46483277099999998,0.2172}
			rgb={0.99939999999999996,0.40860000000000002,0.2172}
			rgb={0.99475999176873464,0.33177297300202935,0.21123096385202059}
			rgb={0.98671295054795893,0.25951834109149341,0.19012239549291934}
			rgb={0.99124588756464194,0.14799417507952672,0.21078892136920357}
			rgb={0.94990303700000001,0.11686717100000001,0.252900603}
			rgb={0.903199533,0.078432949000000002,0.29180038899999999}
			rgb={0.85650000000000004,0.040000000000000001,0.33069999999999999}
			rgb={0.79890262700000003,0.043333450000000003,0.35843429799999998}
			rgb={0.74129942400000004,0.046666699999999998,0.38616694400000001}
			rgb={0.68369999999999997,0.050000000000000003,0.41389999999999999}
}}
\renewcommand{\vec}[1]{%
  \ensuremath{\bm{#1}}%
}

\newcommand{\tnsr}[1]{%
  \ensuremath{\bm{#1}}%
}

\providecommand{\regionint}[3]{%
  \int\limits_{#1}{} {#2} \; {#3}%
}

\DeclarePairedDelimiterX%
  \sDAvg[1]%
    {[}{]_{s}}{#1}

\DeclarePairedDelimiterX%
  \DAvg[1]%
    {[}{]}{#1}

\newif\ifannotated
\annotatedfalse

\newcommand{\annotatedchange}[2]{%
  \ifannotated{\color{gray}{#1}}\fi%
  \ifannotated{\color{magenta}{#2}}\else{#2}\fi%
}

\newcommand{\annotatednew}[1]{%
  \ifannotated{\color{magenta}{#1}}\else{#1}\fi%
}

\AtBeginDocument{%
  \heavyrulewidth=.08em
  \lightrulewidth=.05em
  \cmidrulewidth=.03em
  \belowrulesep=.65ex
  \belowbottomsep=0pt
  \aboverulesep=.4ex
  \abovetopsep=0pt
  \cmidrulesep=\doublerulesep
  \cmidrulekern=.5em
  \defaultaddspace=.5em
}

\makeatletter
\def\@email#1#2{%
 \endgroup
 \patchcmd{\titleblock@produce}
  {\frontmatter@RRAPformat}
  {\frontmatter@RRAPformat{\produce@RRAP{*#1\href{mailto:#2}{#2}}}\frontmatter@RRAPformat}
  {}{}
}%
\makeatother
\begin{document}

\preprint{AIP/PoF}

\title{
Large Eddy Simulation of flow in porous media:\\
analysis of the commutation error of the double-averaged equations
}
\author{W. Sadowski}%
 \email{wojciech.sadowski@rub.de}
\author{M. Sayyari}%
\author{F. di Mare}%
\affiliation{ Chair of Thermal Turbomachines and Aeroengines,\\
Department of Mechanical Engineering, Ruhr University Bochum, 
Universit\"{a}tsstr. 150, Bochum, 44801, Germany
}%
\author{H. Marschall}%
\affiliation{Computational Multiphase Flow, Department of Mathematics,\\
Technical University Darmstadt, 
Alarich-Weiss-Str. 10, Darmstadt, 64287, Germany
}


\date{\today}

\begin{abstract}
The continuum approach employing porous media models is a robust and efficient
solution method in the area of the simulation of fixed-bed reactors. This paper
applies the double-averaging methodology to refine the continuum approach,
opening a way to alleviate its main limitations: space-invariant averaging
volume and inaccurate treatment of the porous/fluid interface. The averaging
operator is recast as a general space-time filter allowing for the analysis of
commutation errors in a classic \emph{Large Eddy Simulation} (LES) formalism. An
explicit filtering framework has been implemented to carry out an \emph{a-posteriori}
evaluation of the unclosed terms appearing in the \emph{Double-Averaged Navier-Stokes}
(DANS) equations, also considering a space-varying filter width. Two resolved
simulations have been performed. First, the flow around a single, stationary
particle has been used to validate derived equations and the filtering
procedure. Second, an LES of the turbulent flow in a channel partly occupied
with a porous medium has been realised and filtered. The commutation error at the
porous-fluid interface has been evaluated and compared to the prediction of two
models. The significance of the commutation error terms is also discussed and
assessed. Finally, the solver for DANS equations has been developed and used to
simulate both of the studied geometries. The magnitude of the error associated
with neglecting the commutation errors has been investigated and an LES
simulation combined with a porous drag model was performed. Very encouraging
results have been obtained indicating that the inaccuracy of the drag closure
overshadows the error related to the commutation of operators.
\end{abstract}

\maketitle

\section{Introduction}

\emph{Commutation error} (CE) terms are the result of the mathematical
derivation of averaged differential equations. Their presence arises from the
commutation of the averaging and differentiation operators. In the framework of
\emph{Large Eddy Simulation} (LES) filtering (see Refs.~\onlinecite{sagaut2006,fureby1997,vasilyev2004}), they
appear when the averaging (or filtering) is inhomogeneous (i.e.\ the averaging
volume changes with space) or when it is applied near the domain boundary (the
averaging volume is extended outside the domain). Exploring both of these
situations is an important issue also in the context of the continuum
description of flows in porous media or particle assemblies. For example,
inhomogeneous filtering may be useful when dealing with porous media with
highly non-uniform pore size distribution. In such conditions, the commutation
error terms can have a measurable contribution, thus, ignoring them may induce
a significant simulation error. The present paper aims to derive and introduce CE
terms as source terms (effectively eliminating, or correcting, the induced
error) in the context of a space-time averaged flow simulation in porous media.

The simulations of reacting gas flows passing over packed beds are
challenging from both the geometrical and computational perspectives. 
Two recent approaches for such flows are:
\emph{particle-resolved Simulation} (PRS), where the geometry of each pellet is
accurately represented (see e.g.~\cite{jurtz2019} or~\cite{dixon2020}),
and \emph{homogenised} modelling, where the packed bed is modelled as a porous
medium~\citep{collazo2012,woudberg2020,mahiques2023}. 
The PRS is a computationally demanding method, due to the meshing requirements
imposed by complex geometries arising inside the packed bed. On top of that,
additional complexity is introduced when the flow is turbulent and needs to be
either resolved or modelled. We refer the interested reader to the works
of~\citeauthor{shams2012}\cite{shams2012, shams2013a, shams2013b, shams2013c,
shams2013d, shams2014a, shams2015} for an exhaustive review of turbulence modelling
approaches for packed beds. 

On the other hand, homogenised modelling usually splits the simulation into
two distinct regions: the freeboard unobstructed by the particles or pallets
and the porous region created from spatial averaging the flow around said
particles~\citep{collazo2012}. The interface between them is often treated 
in a discontinuous way.
In the porous region, the simulation can be coupled with the \emph{Discrete
element method} (DEM) to take into account the interactions between
particles~\citep[see e.g.][]{wiese2016}. The effect of the pallets on the flow
in the homogenised region is described using drag models such as the
Darcy-Forchheimer~\citep{whitaker1996} or Ergun~\citep{ergun1952} equations.

A typical treatment of the boundary between the two regions is to use a
coupling condition ensuring continuity of the flow variables while prescribing
a stress jump at the interface~\citep{ochoatapia1995}. This approach has been
extended to include more complex physics, for example, using averaged
quantities from the bed~\citep{porteiro2009}, including radiation fluxes and
temperature distribution~\citep{rajika2016}, or taking into account the
velocity of the moving interface~\citep{fernando2016}. However, none of the
interface models was entirely successful in accounting accurately for the
influence of slip velocity and turbulent interactions, without tweaking the
empirical model constants. Additionally, a discontinuous approach is difficult
to use in geometries where the interface is hard to define exactly.

An alternative is to describe the interface as a continuous transition between
the two regions and model the flow using a single two-phased set of equations.
Such a set of equations can be obtained through the \emph{Volume Averaging
Theory} (VAT) \citep{whitaker1999}. The \emph{Volume Averaged Navier--Stokes}
(VANS) equations have been a good starting point for a {rigorous} derivation of
Darcy's and Forchheimer's equations, and two-phase porous media flow
models~\citep{whitaker1986a, whitaker1996, whitaker1986b}. In the VAT
framework, unclosed terms describing drag forces and dispersion stresses appear
as a result of the averaging operation. This derivation enables estimating
their values and an in-depth analysis of their effects. The physical
conditions, including the description of the interface \citep[see
e.g.][]{goyeau2003}, can be described by space- and time-varying parameters
like porosity. Ideally, within this approach, the turbulent mixing and other
processes at the interface would be embedded in the turbulence closure.

To extend the VAT to moving beds and turbulent flows, the temporal and spatial
averaging are used sequentially, leading to the \emph{Double Averaged
Navier--Stokes}, or DANS, equations (for an extensive review the reader is
reminded to \citet{lage2002}). \annotatedchange{The main advantage of the DANS equations is that
it describes averaged flow properties in time and space,
allowing to account
for the movement of the bed through the introduction of a space-time porosity,
\(\phi_{VT}\)~\citep{nikora2007a, nikora2013}.}{The main advantage of the DANS equations is that
it describes averaged flow properties in time and space~\citep{nikora2007a,nikora2013}, for example, enabling the analysis of the second-order moments~\citep{papadopoulos2020a}.
The movement of the bed is accounted for through the introduction of a space-time porosity
\(\phi_{VT}\).}
The basis for the definition of
the double-average are the averaging volume and time window, \(V_0\) and
\(T_0\) respectively, chosen so that the macroscopic variations of the flow
parameters  (e.g.\ large-scale velocity fluctuations) are not filtered out. The
underlying assumption is that the characteristic dimension of \(V_0\) will be
much larger than the dimension of small-scale flow parameters (turbulent
eddies, features of porous matrix) but smaller than the characteristic
dimension of the whole domain \citep{gray1975, whitaker1999}. On
the other hand, less consideration was given to the choice of \(T_0\) in
literature. \citet{nikora2013} state that \(T_0\) must be greater that the
turbulent time scale but smaller than the scale of changes in the bed
structure. \citet{vowinckel2017b} followed this approach to evaluate the
double-averaged statistics of the flow over a moving granular assembly. This
means that for a stationary porous structure and flow, an infinite averaging
time window can be used, similar to the standard average used in the
\emph{Reynolds averaged Navier--Stokes} (RANS)~\citep{wilcox1993} modelling.

This similarity between the DANS approach and turbulence modelling also extends to
spatial averaging, which is equivalent to LES filtering. This observation
highlights the other advantage of DANS equations, laying in the ability to
bridge the description of porous flow with classical time- and space-averaging
based turbulence closures. That being said, the filtering scale is usually much
smaller in LES than inside the porous region, therefore, such an approach
requires, at least formally, a different filter size in different parts of the
domain. Having that in mind, deriving such ``combined'' models requires an
in-depth analysis of averaging operators and operations used in the derivation
of the equations, both from a mathematical and numerical standpoint.
Initial development of the space-varying averaging in the context
of VAT was investigated, for example, by \citet{gray1982}, who assumed a
spherical averaging volume and derived modified averaging theorems, \emph{de
facto} correcting the commutation errors of the averaging and differentiation
operators. 

Investigation of commutation errors is also important for explicit averaging
the results from particle-resolved simulations. While the usage of the PRS technique is
getting more popular, filtered data can be used to develop more accurate drag
and turbulence models, potentially free of empirical correlations. One example
of such an approach is the evaluation of the momentum balance in the DANS
equations for flows over moving river beds done by \citet{vowinckel2017b} or
the analysis of the budget of turbulence kinetic energy
\citep{papadopoulos2020b} based on the same data. However, no additional
consideration was given to the treatment of commutation errors in both studies.
A more thorough investigation of the effects of neglecting CEs was presented
by~\citet{iovieno2003variable}. The authors developed a procedure of
approximating commutation errors numerically and compensating for their
effects by including them as source terms. They investigated the effect both
numerically and analytically, in both cases, it was shown that neglecting CEs
induces errors, which also influence the flow in regions where the filter
length is constant. The authors suggested that the filtering procedure is also
applicable to explicit filtering~\citep{iovieno2003variable}. We use a similar
technique to arrive at DANS equations with commutation error treatment as
source terms, with a focus on the analytic form of these source terms.
\citet{klein2020} studied the effects of commutation error induced by a filter
with a size set by a stretched grid, with a fixed stretching factor. They
applied the filtering to isotropic turbulence generated as initial data, and
to Direct Numerical Simualtion (DNS) results of a turbulent channel flow. They used the differential
approximation of the filter and a model based on the scale-similarity
hypothesis, to correct the CEs, where in all cases, their modelling approach
yielded lower errors than when the commutation errors were neglected. We
compare their formulation to CEs computed from an LES of a channel flow over a
porous matrix.

The main focus of the present work is to extend the approach developed by
\citet{nikora2013} to achieve the following goals:
\begin{enumerate}
  \item Recasting the definition of double averaging operation as filtering,
    determining the requirements for the filtering operator and using it to
    rederive the DANS equations. Moreover, including a description of
    commutation errors, allowing for inhomogeneous filtering and accurate
    treatment of filtering near boundaries.
  \item Numerical verification of all derived error terms by \emph{a-priori} testing on a
    simplified resolved particle simulation.
  \item \emph{A-posteriori} analysis of the non-uniform filtering induced commutation error in a fully 
    developed turbulent flow over a porous medium.
  \item Initial development of a solver for double-averaged equations, 
    capable of performing scale-resolving simulation coupled with porous
    drag closure.
\end{enumerate}

This work is arranged as follows. In Section~\ref{s:avg_approach}, we detail
the mathematical model used in this paper. We introduce the filtering operator based on
superficial averaging and compute the commutation errors arising from
filtering the space and time derivatives, in Sections~\ref{s:comm_err}
and~\ref{s:cat_comm} respectively. Next, in Section~\ref{s:dans}, we
apply the filtering to Navier--Stokes equations and describe the commutation
errors arising from this operation. Following that, in
Section~\ref{s:numerics}, we specify the numerical approaches used in this
work, and we lay the groundwork to develop a solver for the DANS equations. In
Section~\ref{s:test-cases}, we detail the test cases used to demonstrate the
effects of commutation error terms. Further, in Section~\ref{s:results}, we
discuss the results from the simulations. Lastly, we conclude and
provide an outlook in Section~\ref{s:conclusions}.

\section{Methodology}\label{s:avg_approach}

\begin{figure}
  \includegraphics[width=\linewidth]{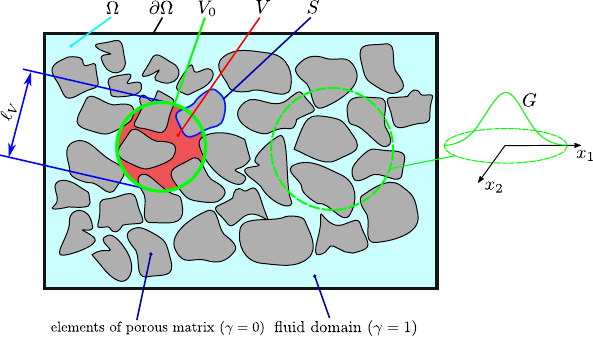}
  \caption{\label{fig:schamtic}Schematic description of the porous domain with definitions related
  to volume averaging/filtering: \(\Omega\) - the computational domain, \(\partial
  \Omega\) - the boundary of the computational domain, \(S\) - the interface between
  porous matrix elements and fluid, \(V_0\) - averaging volume, \(V\) - the volume
  occupied by the fluid inside the averaging volume, \(\ell_V\) - dimension of the
  averaging volume/cutoff length scale of the filtering kernel, \(G\) - filtering
  kernel.}
\end{figure}

Various types of averaging operations are known in the context of multiphase
and porous media modelling. The most fundamental distinction must be made
between superficial and intrinsic average~\citep{faghri2020}. The first one is
taken in the whole averaging domain, whereas in the latter, the integration
domain is restricted to a subset of the space (in case of double-averaging also
time) in which fluid is present (see Figure~\ref{fig:schamtic}). The
double-average can be defined as a consecutive time-space average, averaging in
time than in space, the consecutive space-time average or general space-time
average where both integrals are applied simultaneously. Intrinsic versions of
these averages are not equivalent to each other, however, the superficial
space-time average and its consecutive counterparts are all equal, as shown
by~\citet{nikora2013}. Owing to that, we can restrict our investigation to the
analysis of space-time superficial average without any loss of generality.



\subsection{Definition of the filtering operator}\label{s:comm_err}
A classical description of the double-averaging technique is presented for
completeness in Appendix~\ref{s:classical_averaging}. Importantly, the main
assumption behind the derivation of the averaging theorems is the fact that
\emph{the averaging volume remains space-invariant}. To generalise it beyond
this constraint and derive the analytical description of the commutation error
source terms, we define the filtering operator and write a superficial average
based on its kernel. 

A space-time filtering kernel \(G\) can be defined as a product of time-based
kernel \(H(t, T_0)\) and a spatial one \(\widehat{G}(\vec{x}, \ell_V)\), where
\(\ell_V\) denotes the characteristic size of
\(\widehat{G}\)~\citep{sagaut2006}. In the definition of \(G\), both \(H\) and
\(\widehat{G}\) can be arbitrary kernels, provided that they satisfy the
following properties. First, \(G(\vec{x}, t, \ell_V, T_0) = H\widehat{G}\) has
to be linear and conserve constants. Second, for {consistency}, both kernels
must approach Dirac \(\delta\) function in the limit of vanishing filtering
width. The superficial average can be written as
\begin{equation}
  [\psi]_s(\vec{x}, t) = G \star (\gamma\psi) =
   \displaystyle\int\limits_{-\infty}^{\infty}
      \int\limits_{\Omega}
      G(\vec{x} - \vec{\xi}, t - \tau, \ell_V, T_0)
      \gamma(\vec{\xi}, \tau)\psi(\vec{\xi}, \tau)
      \;\text{d} \vec{\xi}\;\text{d} \tau 
    \label{eq:superficial_convolution}
\end{equation}
where \(\gamma\) is a clipping or phase indicator function
\citep{nikora2013, breugem2006} and is equal to 1 in the fluid and 
0 in the solid phase. This definition leads to the same formula for
space-time porosity as described in Appendix~\ref{s:classical_averaging}, i.e.\
 \(\phi_{VT} = \sDAvg{1}\). Similarly, the superficial average and the
 space-time porosity \(\phi_{VT}\) are related to the intrinsic
 average denoted with \(\DAvg{\cdot}\) 
\begin{equation}\label{eq::sup_int_relation}
  \sDAvg{\psi} = \phi_{VT}\DAvg{\psi}.
\end{equation}

 This allows us
to use the double-filtering framework as a generalisation of both spatially and
temporally averaged Navier--Stokes equations. When \(\widehat{G}\) is assumed
to be a spherical top-hat filter and \(T_0 = 0\), the standard volume averaging
approach is recovered. Similarly, when \(\ell_V = 0\) and \(H\) is a top-hat
filter, the (U)RANS-like average is obtained.


\subsection{Derivation of averaging theorems with commutation errors}
\label{s:cat_comm}

Methods used for deriving the averaging theorems are similar to the formulas
describing commutation errors in non-homogeneously filtered
LES~\citep[see][]{fureby1997, sagaut2006}. Merging the
approach\cite{fureby1997} and our definition of filtering given by Eq.
\eqref{eq:superficial_convolution}, space-time filtering can be extended to
non-homogeneous filters. For this, we introduce a few assumptions guiding this
derivation. For simplicity, we consider the cut-off length \(\ell_V\) and
time-averaging window \(T_0\) to be dependent only on space, \(\ell_V =
\ell_V(\vec{x})\), and time, \(T_0 = T_0(t)\), respectively. Additionally, the
extension of the domain occupied by the fluid can change with time, i.e.\
\(\Omega = \Omega(t)\). We define this change by a function \(D(\vec{x},
t)\in\mathds{R}^3\times\mathds{T}\), equal to 1 in the computational domain
\(\Omega(t)\) and 0 otherwise (similarly to \(\gamma\)). 

The first step in evaluating the commutation error will be taking the
derivative of the convolution integral~\eqref{eq:superficial_convolution} with
respect to either \(x_i\) or \(t\), which we denote with \(s\) as a
placeholder. For simplicity, we omit function dependencies since all functions
depend on $s$.

\begin{eqnarray}
  \lefteqn{
  \dpd{\sDAvg{\psi}}{s}
    = \dpd{}{s}\regionint{\Omega(t) \times\mathds{T}}{G \gamma \psi}{\dif \vec{\xi} \dif \tau} 
    = \dpd{}{s}\regionint{\mathds{R} \times\mathds{T}}{G \gamma \psi D
      }{\dif \vec{\xi} \dif \tau} }\nonumber\\
    &=
    \displaystyle\regionint{\mathds{R} \times\mathds{T}}{%
      \dpd{G \gamma \psi}{s}D
    }{\dif \vec{\xi} \dif \tau}
    + \regionint{\mathds{R} \times\mathds{T}}{%
        G \gamma \psi \dpd{D}{s}
      }{\dif \vec{\xi} \dif \tau}\label{eq::comm_long_1}\\
    &= \underbrace{\regionint{\Omega(t) \times\mathds{T}}{%
        \dpd{G \gamma \psi}{s}
      }{\dif \vec{\xi} \dif \tau}}_{=I_1}
    + \underbrace{\regionint{\mathds{R} \times\mathds{T}}{%
        G \gamma \psi \dpd{D}{s}
  }{\dif \vec{\xi} \dif \tau}}_{=I_2}\nonumber.
\end{eqnarray}

The term \(I_2\) in Eq.~\eqref{eq::comm_long_1} is related to the error arising 
when filtering in the bounded domain. Term \(I_1\) can be decomposed into three 
integrals, after employing the product
rule and relations from \ref{s:app_drel} (assuming here that \(s=x_i\)).
\begin{eqnarray}
  I_1\Big|_{s=x_i} 
  &=& \dpd{\ell_V}{x_i} \regionint{\Omega(t) \times \mathds{T}}
    {\dpd{G}{\ell_V} \gamma \psi }{\dif \vec{\xi} \dif \tau} 
    + \regionint{\Omega(t) \times \mathds{T}}
    {G \gamma \dpd{\psi}{x_i} }{\dif \vec{\xi} \dif \tau}
  + \regionint{\Omega(t) \times \mathds{T}}
  {G  n_i \delta(\vec{x} - \vec{\xi} -  \vec{x}_S) \psi }{\dif \vec{\xi} \dif \tau}
       \nonumber\\
  &=& \dpd{\ell_V}{x_i} \left\{ \dpd{G}{\ell_V} \star (\gamma \psi) \right\}
  + \regionint{S(t) \times \mathds{T}} {G  n_i \psi }{\dif \vec{\xi} \dif \tau}
  + G \star \left(\gamma \dpd{\psi}{x_i}\right).
\end{eqnarray}
The derivatives of \(D\) and \(\gamma\) can be treated in the same way. This
leads to the following simplification of \(I_2\) term:
\begin{equation}
  I_2\Big|_{s=x_i} 
  = \regionint{\partial \Omega(t) \times \mathds{T}} {G \gamma \psi n^{\Omega}_i }{\dif \vec{\xi} \dif \tau}
\end{equation}
Here, \(\vec{n}^{\Omega}\) is the inward normal vector of the domain boundary.
After rearranging the terms, the final formula relating the time derivative
of the filtered variable and the filtered derivative reads:
\begin{equation}
  \sDAvg*{\dpd{\psi}{t}} 
   = \dpd{\sDAvg{\psi} }{t} 
   + \regionint{S(t) \times \mathds{T}} {G  w_i n_i \psi }{\dif \vec{\xi} \dif \tau}
   - \dpd{T_0}{t} \left\{ \dpd{G}{T_0} \star (\gamma \psi) \right\}
   + \regionint{\partial \Omega(t) \times \mathds{T}} 
    {G \gamma \psi w^{\Omega}_i n^{\Omega}_i }{\dif \vec{\xi} \dif \tau}.\label{eq::comm_t}
\end{equation}
Similarly, for the space derivative:
\begin{equation}
  \sDAvg*{\dpd{\psi}{x_i}} 
  = \dpd{\sDAvg{\psi}}{x_i} 
  - \regionint{S(t) \times \mathds{T}} {G  n_i \psi }{\dif \vec{\xi} \dif \tau} 
  - \dpd{\ell_V}{x_i} \left\{ \dpd{G}{\ell_V} \star (\gamma \psi) \right\}
  - \regionint{\partial \Omega(t) \times \mathds{T}} {G \gamma \psi n^{\Omega}_i }{\dif \vec{\xi} \dif \tau}.
\label{eq::comm_x}
\end{equation}

Inspection of the resulting relations leads to the following remarks:
\begin{enumerate}[label=(\roman*)]
  \item The integrals over the interface \(S\) are identical to the ones
    presented in equations \eqref{eq::da_theorem_t} and
    \eqref{eq::da_theorem_x} (the scaling \(1/V_0\) is implicitly included in
    the filter kernel \(G\)).
  \item In the limits of homogeneous filtering, the derivatives of \(\ell_V\)
    over space and \(T_0\) over time vanish. Similarly, on unbounded domains, the
    boundary terms in equations~\eqref{eq::comm_t} and
    \eqref{eq::comm_x} also vanish. Then, the equations reduce to equations
    \eqref{eq::da_theorem_t} and \eqref{eq::da_theorem_x} respectively. This
    shows that the derived equations are a general form of averaging theorems.
  \item The terms containing the derivatives of \(G\) describe the error
    arising when filtering volume is not homogeneous in time and space. This
    error might be modelled by including those terms in the DANS system. The
    ensuing error definitions are the generalisation of integrals derived by
    \citet{gray1982} for spherical averaging volume.
\end{enumerate}
 
The surface integrals over \(\partial \Omega(t)\) represent the error arising
from filtering near the domain boundary. It should be noted that these have the
same form as surface integrals in averaging theorems. Therefore, the boundary
can be divided into two parts, \(\partial \Omega = \partial \Omega^P  +
\partial \Omega^C\) where \( \partial \Omega^P \) is the part directly adjacent
to porous matrix elements and \(\partial \Omega^C\) limits the free fluid
(boundary surface where \(\phi_{VT} = 1\)). The term \(\partial \Omega^P\), which
can be thought of as a rough boundary, can be incorporated into the unclosed
source terms (e.g.\ by extending \(\gamma\) outside the domain assuming
\(\gamma =0\)) and the commutation error related to this part of the boundary
could be fully modelled as a drag force. The surface terms are a generalisation
of similar relationships obtained for LES filtering \citep{fureby1997}. In the
work of \citet{fureby1997}, however, an inconsistent choice of the
direction of boundary normal vectors with the sign in the expression for
\(\partial D / \partial x_i\) was made, which has been corrected in the present
study. 

%

\subsection{Commutation error terms in the DANS equations}\label{s:dans}
Utilizing the averaging theorems, we analyse the necessary steps to
reduce the 
Navier--Stokes equations using the double-averaging approach discussed in the
previous section. We consider the system of the Navier--Stokes equations,
assuming isochoric, isothermal and low-mach flows (material properties like
density or viscosity remain constant):
\begin{equation}\label{eq:ins}
  \left\{
  \begin{array}{rl}
	  \dpd{\rho}{t} + \dpd{\rho u_j}{x_j} &= 0\;,\\
    \dpd{\rho u_i}{t} + \dpd{\rho u_i u_j}{x_j} 
    &= \dpd{\sigma_{ij}}{x_j}  + \rho f_i\,,
  \end{array}
\right.
\end{equation}
with \(\rho\), \(\vec{u}\) and \(\vec{f}\) denoting respectively density,
velocity and body forces. The stress tensor is given as \(\tnsr{\sigma} = -p
\tnsr{I} + \mu (\nabla \vec{u} + \nabla\vec{u}^T)\). The density is kept inside
the derivatives on purpose, as it facilitates the derivation below. In the
above equations and throughout the whole paper, the repeated indices imply
summation. Application of theorems given by Eq. \eqref{eq::da_theorem_t} and
\eqref{eq::da_theorem_x}, leads to the Double--Averaged Navier--Stokes system,
as defined by \citet{nikora2013}:
\begin{subequations}\label{eq::da_intrinsic}
\begin{eqnarray}
  \dpd{\phi_{VT}}{t} + \dpd{ \phi_{VT} \DAvg{u_i} }{x_i} &=& 0,\\
	  \rho\left(
      \dpd{\phi_{VT}\DAvg{u_i}}{t} + \dpd{\phi_{VT}\DAvg{u_i}\DAvg{u_j}}{x_j}
  \right)  &=&
    \dpd{\phi_{VT} \DAvg{\sigma_{ij} }}{x_j}
       + \dpd{\rho\phi_{VT} \tau_{ij} }{x_j}
     + \rho\phi_{VT}\DAvg{f_i} + \rho F_i,
\end{eqnarray}
\end{subequations}
where \(\tau_{ij} = \DAvg{u_i}\DAvg{u_j} - \DAvg{u_i u_j}\) is the sub-filter
stress and \(\vec{F}\) denotes porous-induced drag force. The derivation of the above system is provided in Appendix~\ref{s:dans_app}.


Instead, to obtain the commutation error source terms for the DANS equations, we define
the filtering with Eq. \eqref{eq:superficial_convolution} and apply the
generalisations of the averaging theorems, Equations \eqref{eq::comm_t} and
\eqref{eq::comm_x}, to the Navier-Stokes system. With this in mind, 
drag force can be redefined as 
\begin{equation*}
  F_i = - 
\regionint{S(t) \times \mathds{T}} {G\sigma_{ij} n_j}{\dif \vec{\xi} \dif \tau}
\end{equation*}
and a new DANS system can be formulated. For the
momentum equation, the following terms are thus obtained, related to changes of
\(T_0\) \eqref{eq::FparT}, change of \(\ell_V\) \eqref{eq::FparL} and filtering
near the boundaries \eqref{eq::Fb}
\begin{subequations}
  \begin{eqnarray}
    F^{T_0}_i &=& \dpd{T_0}{t} \left( \dpd{G}{T_0} \star (\gamma u_i)\right),\label{eq::FparT}\\
    F^{\ell_V}_i &=& 
  \dpd{\ell_V}{x_j} \left( \dpd{G}{\ell_V} \star (\gamma u_i u_j - \gamma\sigma_{ij}) \right),
  \label{eq::FparL}\\
    F^\Omega_i &=&
  \regionint{\partial \Omega(t) \times \mathds{T}} 
  {G \gamma( -u_i w^{\Omega}_j  +  u_i u_j - \sigma_{ij}) n^{\Omega}_j }{\dif \vec{\xi} \dif \tau}.\label{eq::Fb}
  \end{eqnarray}
\end{subequations}
Analogous terms appear for the continuity equation
\begin{subequations}
  \begin{eqnarray}
    \Lambda^{T_0} &=& \dpd{T_0}{t} \left( \dpd{G}{T_0} \star \gamma \right)\label{eq::LparT},\\
    \Lambda^{\ell_V} &=& \dpd{\ell_V}{x_j} \left( \dpd{G}{\ell_V} \star (\gamma u_j) \right)\label{eq::LparL},\\
    \Lambda^\Omega &=& \regionint{\partial \Omega(t) \times \mathds{T}} 
  {G \gamma( - w^{\Omega}_i  +  u_i ) n^{\Omega}_i }{\dif \vec{\xi} \dif \tau}\label{eq::Lb}.
  \end{eqnarray}
\end{subequations}

Considering that \(F_i^c = F^{T_0}_i + F^{\ell_V}_i + F^\Omega_i\)
and \(\Lambda^c = \Lambda^{T_0} + \Lambda^{\ell_V} + \Lambda^\Omega\), the set of
corrected double-filtered equations has the following form:
\begin{subequations}\label{eq::da_intrinsic_comm_terms}
  \begin{eqnarray}
    \dpd{\phi_{VT}}{t} + \dpd{ \phi_{VT} \DAvg{u_i} }{x_i} &=& \Lambda^c,\\
	  \rho \left( 
      \dpd{\phi_{VT}\DAvg{u_i}}{t} + \dpd{\phi_{VT}\DAvg{u_i}\DAvg{u_j}}{x_j} 
  \right)  &=&
    \dpd{\phi_{VT} \DAvg{\sigma_{ij} }}{x_j}
    + \dpd{\phi_{VT} \tau_{ij} }{x_j}
  + \rho\left(\phi_{VT}\DAvg{f_i} + F_i + F^c_i\right).
  \end{eqnarray}
\end{subequations}


\subsection{Commutation error terms of the \annotatedchange{avaeraged}{averaged} stress tensor}

The equations above are very similar to the Navier--Stokes system. However, in
order to implement them in the context of a numerical framework (e.g.\ finite
volume formulation), the double-averaged stress tensor \(
\DAvg{\tnsr{\sigma}}\) needs to be described as a function of averaged
velocity. To this aim, it is split into pressure and viscous part
\begin{equation}
  \dpd{\phi_{VT} \DAvg{\sigma_{ij} }}{x_j} = -\dpd{\phi_{VT}\DAvg{p}}{x_i} 
+ \dpd{\sDAvg{D_{ij} }}{x_j}.
\end{equation}
Using Equation \eqref{eq::comm_x} on the viscous stress tensor
\(\sDAvg{\tnsr{D}}\) moves the derivatives onto the velocity, generating
new unclosed and commutation errors terms:
\begin{equation}\label{eq:dij_true}
    \sDAvg{D_{ij}} = 
    \mu \left( \dpd{\phi_{VT}\DAvg{u_i}}{x_j} + \dpd{\phi_{VT}\DAvg{u_j}}{x_i} \right)
    - D^{\ell_V}_{ij} - D^{S}_{ij} - D^{\Omega}_{ij}
\end{equation}
where tensors \(\tnsr{D}^{\ell_V}\), \(\tnsr{D}^{S}\), \(\tnsr{D}^{\Omega}\) are:

\begin{subequations}
  \begin{eqnarray}
    D^{\ell_V}_{ij} &=& \mu \left\{ 
    \dpd{\ell_V}{x_j}\left( \dpd{G}{\ell_V}\star(\gamma u_i) \right)   
    + \dpd{\ell_V}{x_i}\left( \dpd{G}{\ell_V}\star(\gamma u_j) \right)
  \right\}\label{eq::DparL},\\
      D^{S}_{ij} &=& \mu 
  \regionint{S(t) \times \mathds{T}} 
  {G (u_i n_j + u_j n_i) }{\dif \vec{\xi} \dif \tau}
  ,\\
  D^{\Omega}_{ij} &=& \mu
  \regionint{\partial \Omega(t) \times \mathds{T}} 
  {G\gamma (u_i n^\Omega_j + u_j n^\Omega_i) }{\dif \vec{\xi} \dif \tau}\label{eq::Db}.
  \end{eqnarray}
\end{subequations}

Inserting the above into the momentum equation and writing 
\(\tnsr{D}^c = \tnsr{D}^{\ell_V} + \tnsr{D}^{\Omega}\), results in:
\begin{widetext}
\begin{equation}
  \begin{split}
    &\rho \left( \dpd{\phi_{VT}\DAvg{u_i}}{t} +
    \dpd{\phi_{VT}\DAvg{u_i}\DAvg{u_j}}{x_j} \right)  
    = 
    - \dpd{\phi_{VT}\DAvg{p}}{x_i} 
    + \dpd{}{x_j}\left\{ \mu \left( \dpd{\phi_{VT}\DAvg{u_i}}{x_j} +
      \dpd{\phi_{VT}\DAvg{u_j}}{x_i} \right) \right\} \\
    &+ \dpd{\phi_{VT} \tau_{ij} }{x_j} 
  + \rho\left(\phi_{VT}\DAvg{f_i} + F_i + F^c_i\right)
  - \dpd{D^{S}_{ij}}{x_j}
  - \dpd{D^{c}_{ij}}{x_j},
  \end{split}
\end{equation}
\end{widetext}
which is the form with all of the commutation errors included.

For the cases presented in this work, the obstacles building up the porous
medium are stationary, hence the porosity field is constant in time. In that
case, we may simplify the momentum and continuity equation. To make the notation
clearer, we define \(\phi\equiv\phi_{VT}\) and given the isochoric conditions,
we incorporate the density into the pressure variable. Due to the no-slip
condition at \(S\), \(\tnsr{D}^S=\tnsr{0}\). We also decompose the pressure
gradient and viscous stress tensor:
\begin{equation}
  \dpd{\phi\DAvg{p}}{x_i} = \phi\dpd{\DAvg{p}}{x_i} + \DAvg{p}\dpd{\phi}{x_i},
  \label{eq:press_gradient_decomp}
\end{equation}
\begin{equation}
  \nu \left( \dpd{\phi\DAvg{u_i}}{x_j} +
      \dpd{\phi\DAvg{u_j}}{x_i} \right)
  =
  \phi\nu \left( \dpd{\DAvg{u_i}}{x_j} +
      \dpd{\DAvg{u_j}}{x_i} \right)
  +
  \nu \left( \dpd{\phi}{x_j}\DAvg{u_i} + 
  \dpd{\phi}{x_i}\DAvg{u_j} \right).
  \label{eq:vis_sttress_decomp}
\end{equation}
The term \(\tilde{\vec{F}}\) groups
all source terms present in momentum equations, apart from the drag force
\begin{equation}
  \tilde{F}_i 
  = \phi\DAvg{f_i} + F^c_i - \dpd{ D^c_{ij}}{x_j} 
  +\dpd{}{x_j}\left\{\nu \left( \dpd{\phi}{x_j}\DAvg{u_i} + 
    \dpd{\phi}{x_i}\DAvg{u_j}\right)
  \right\},
  \label{eq:mom_source}
\end{equation}
resulting in a following DANS system
\begin{subequations}\label{eq:dans_steady_por}
  \begin{eqnarray}
    \dpd{ \phi \DAvg{u_i} }{x_i} &=&\Lambda^c\\
    \phi\dpd{\DAvg{u_i}}{t}+
    \dpd{\phi\DAvg{u_i}\DAvg{u_j}}{x_j} &=& \dpd{\phi\DAvg{p}}{x_i}
    + \dpd{}{x_j}\left\{ \phi\nu \left( \dpd{\DAvg{u_i}}{x_j} +
      \dpd{\DAvg{u_j}}{x_i} \right) \right\}
      + \dpd{\phi \tau_{ij} }{x_j} 
  + F_i
  + \tilde{F}_i.
  \end{eqnarray}
\end{subequations}

%
%

\section{Numerical method}\label{s:numerics}

In this section, we describe the numerical methods and their setup for the test
cases studied in the paper. Additionally, we discuss the schemes used in PRSs, the choice of
filter, and the DANS solver.

\subsection{Explicit filtering of PRS}
The fields computed in the PRSs were filtered to arrive at the double-averaged
quantities. The filtering is done using the postprocessing utility described
and validated in {our previous work}~\cite{sadowski2023}. To ensure consistent
results across different spatial kernels, each kernel must be scaled properly.
This is achieved using the condition 
\begin{equation}\label{eq:kernel_size}
  \ell_V^N = \regionint{\Omega}{
    \dfrac{\widehat{G}(\vec{x}, \ell_V)}{\widehat{G}(\vec{0}, \ell_V)}
  }{\dif \vec{x}}, 
\end{equation}
where \(N\) is the dimensionality of the considered domain. We also restrict
ourselves to \(C^0\) kernels, as any discontinuities would be inaccurately
reproduced on the general unstructured mesh, introducing an errorin to the
filtering process (e.g.~the filter will not preserve constants).

In this work, two filters were considered, a Gaussian kernel, and a so-called
Cellular kernel. The Gaussian filter is given by
\begin{equation}
  \widehat{G}_G = \frac{1}{\ell_V^{N/2}}\exp \left(\pi x_i x_i / \ell_V^2 \right),
  \label{eq:gaussian}
\end{equation}
with \(N\) denoting the dimensionality of the domain. Whereas the Cellular
kernel, introduced in~\cite{quintard1994}, is defined in one spatial dimension
as
\begin{equation}
  \widehat{G}_C = \frac{\ell_V - |x|}{\ell_V^2}.
  \label{eq:celular}
\end{equation}
For a spatially periodic porous medium, the cellular filter can be generalized
to three-dimensions by multiplying the filters representing each dimension~\cite{quintard1994, breugem2005}. 

To evaluate the commutation errors related to the change of \(\ell_V\), the
derivative \(\partial \widehat{G} /\partial \ell_V \) has to be evaluated. In
the present work, only Gaussian kernel will be considered in this context and
its derivative is computed as 
\begin{equation}
  \dpd{\widehat{G}_G}{\ell_V} = -\frac{1}{2\ell_V^3}\left(N\ell_V^2-4\pi x_i
  x_i\right)
  \widehat{G}_G.
  \label{eq:gaussian_derivative}
\end{equation}

To speed up both filtering and computing the distributions of \(\widehat{G}\),
the kernel distribution can be clipped to 0 after a certain distance \(d\) from
the center of filtering molecule. We restrict this clipping 
such that the error of filtering operation introduced by clipping is less than
\(1\%\), i.e.\ the integral of the clipped kernel is at least equal to \(0.99\).
A similar procedure can be followed when using
\(\partial \widehat{G} /\partial\ell_V\). However, here the measurement of error 
is less straightforward as the integral of the derivative of the kernel is 
by definition equal to \(0\):
\begin{equation}
  \int\limits_\Omega \pd{\widehat{G}}{\ell_V} \dif \vec{x} = 
  \pd{}{\ell_V} \int\limits_\Omega \widehat{G} \dif \vec{x}= 
  \pd{1}{\ell_V} = 0.
  \label{eq:kernel_deriv_0}
\end{equation}
Therefore, a sensitivity study was conducted on the influence of the clipping
distance for the accuracy of computed quantities based on the filtering of
results from G1 test case (see \ref{s:test-cases} for description of the
geometry and setup). It is presented in~\ref{s:sens_study}. Both Gaussian
kernel and its derivative have been clipped at \(d/\ell_V=1.8\).


\subsection{Modelling and approximation of the commutation errors}
\label{s:models}

Filtering operation with any well defined kernel can be approximated by solving
following differential equation \cite{sagaut2006}:
\begin{equation}
  \langle \psi \rangle^G - \frac{\alpha_2}{2}\dpd[2]{\langle\psi\rangle^G}{x}
  \approx \psi.
  \label{eq:diff_op_kernel}
\end{equation}
In the above equation we denote an intrinsically filtered quantity with the
brackets \(\langle \cdot \rangle^G \). For simplicity we assume that the
filtering occurs only in space and filter is only one dimensional, i.e.\ \(
G=G(\ell_V, x)\). Additionally, \(\alpha_2\), denotes second moment of the
chosen kernel, given as:
\begin{equation}
  \alpha_2 = \regionint{\Omega}{\xi^2 G(\ell_V, \xi)}{\dif\;\xi}.
  \label{eq:moment_two}
\end{equation}
\annotatednew{Differentiating~{\eqref{eq:diff_op_kernel}} w.r.t.\ filter width}, an approximation for the convolution
product \( (\partial G / \partial \ell_V ) \star \psi\) can be derived:
\begin{equation}
  \frac{\partial G}{\partial \ell_V} \star \psi 
  \approx \annotatedchange{2\ell_V}{\dfrac12\dpd{\alpha_2}{\ell_V}}\dpd[2]{\langle \psi\rangle^G}{x}.
\end{equation}
\annotatednew{In our case, for both kernels~{\eqref{eq:gaussian}}
and~{\eqref{eq:celular}}, {\(\alpha_2\)} is equal to {\(\ell^2_V/6\)}:}
\begin{equation}\label{eq:error_approx}
  \annotatednew{%
    \frac{\partial G}{\partial \ell_V} \star \psi 
    \approx \dfrac{\ell_V}{6}\dpd[2]{\langle \psi\rangle^G}{x}.
  }
\end{equation}

\citet{klein2020} have reported that the above approximation performs quite
well based on the tests on the DNS data. They have also attempted modeling of
the error using the scale similarity hypothesis \cite{sagaut2006}. The
commutation error term can be expressed as
\begin{equation}
  \dpd{\ell_V}{x}\left(\dpd{G}{\ell_V} \star \psi \right)
  \approx C \tau_{G}(\partial_x, \langle\psi\rangle_G)
   = C \left( 
  \left\langle \dpd{\langle\psi\rangle^G}{x} \right\rangle^G 
- \dpd{\langle \langle \psi \rangle^G \rangle^G}{x}
  \right),
  \label{eq:ss_model}
\end{equation}
where \(\tau_{G}(\partial_x, \langle\psi\rangle_G)\) denotes \emph{Generalised Central Moment} (GCM) associated
with \(\langle \partial \langle \psi \rangle^G / \partial x\rangle_G \). The
coefficient \(C\) is of the order of one. Alternatively, a model by
\citet{bardino1983} can be used to determine \(C\) dynamically, using two filter levels. 
Given a second filtering
operation, denoted here with \(\langle \cdot \rangle^F\), which corresponds to
a kernel \(F=G(2\ell_V, x)\), GCM can also be defined for a quantity filtered
with both kernels (described as \(\langle\psi\rangle^{GF}\equiv \langle \langle
\psi\rangle^G\rangle^F\) for brevity):
\begin{equation}
  \tau_{GF}(\partial_x, \langle \psi \rangle^{GF}) = 
   \left\langle \dpd{\langle\psi\rangle^{GF}}{x} \right\rangle^{GF} 
  - \dpd{\langle \langle \psi \rangle^{GF} \rangle^{GF}}{x}, 
  \label{eq:tau_fg}
\end{equation}
and similarly
\begin{equation}
  \tau_F(\partial_x, \langle \psi \rangle^G) = 
   \left\langle \dpd{\langle\psi\rangle^{G}}{x} \right\rangle^{F} 
  - \dpd{\langle \langle \psi \rangle^{G} \rangle^{F}}{x}. 
  \label{eq:tau_g}
\end{equation}
Finally, value of \(C\) can be computed as
\begin{equation}
  C = \frac{\tau_F(\partial_x, \langle \psi \rangle^G)I}{I^2}
  ,\quad I = \tau_{GF}(\partial_x, \langle \psi \rangle_{GF})- \langle \tau_G(\partial_x, \langle \psi\rangle_G)\rangle^F.
  \label{eq:C_ss_coeff}
\end{equation}

\subsection{PIMPLE algorithm for DANS equations}
\label{sec:dans_solver}

As mentioned before, considered test cases, will be simulated using
filtered equations. Since the solid elements are stationary for both
configurations, we can assume that porosity does not depend on time. 
With this, we restrict our simulation to the system~\eqref{eq:dans_steady_por}.

Focusing first on the unsteady simulation of the channel and following
\citet{breugem2005}, we assume following closure for the drag term 
\begin{equation}
  F_i 
  \approx -\dpd{\phi}{x_j}\left\{ 
    - \DAvg{p}\delta_{ij} 
    + \nu \left(\dpd{\DAvg{u_i}}{x_j} + \dpd{\DAvg{u_j}}{x_i}\right)
  \right\}
  -\nu \mathcal{K}^{-1}\left(1 + \mathcal{F}\right)\phi^2\DAvg{u_i}.
  \label{eq:drag_closure}
\end{equation}
Where the first term on the right-hand side is meant to simplify the momentum
equation, according to the estimations by \citet{quintard1994} and the last
defines the standard Darcy-Forchheimer drag model~\cite{whitaker1999}.

Assuming that the porous medium has the form of an array of cubes, the model 
coefficients can be computed in the following way~\citet{irmay1965}.
Considering that 
\(d_p\) denotes particle size, here taken to be equal to the dimension of the
cube, the inverse of permeability \(\mathcal{K}\) is given as 
\begin{equation}
  d_p^2 \mathcal{K}^{-1} 
  = C_K
  \dfrac{ 1-\phi }{ {\left(1 - (1-\phi)^{1/3}\right)}^3
  \left(1 + (1-\phi)^{1/3}\right)}.
  \label{eq:permeability}
\end{equation}
The Forcheimer coefficient \(\mathcal{F}\) can be computed in the following manner:
\begin{equation}
  d_p^2\mathcal{F}\mathcal{K}^{-1} 
  = C_F \left( \dfrac{1-\phi}{\phi}\right)
  \dfrac{\phi\left|\DAvg{\vec{u}}\right| d_p}{\nu}.
  \label{eq:forcheimer}
\end{equation}
\citet{breugem2004} calibrated the model coefficients for a given geometry,
resulting in values \(C_K\approx11.4\) and \(C_F\approx0.4\). 
\annotatednew{It is important to note that the choice of the drag model has a big influence
on the overall accuracy of the computation. Chosen model is a good representation of an isotropic
and fully regular porous matrix as chosen for this analysis, however, when a more complex geometry 
would be adopted, the model should account for more sophisticated physical effects,
e.g.~include the influence of non-isotropy.}
Substituting the
Equation \eqref{eq:drag_closure} into the Equation \eqref{eq:dans_steady_por},
leads to a modified momentum equation given by
\begin{eqnarray}
  \phi\dpd{\DAvg{u_i}}{t}+
    \dpd{\phi\DAvg{u_i}\DAvg{u_j}}{x_j} 
    &=& 
    \phi\dpd{\DAvg{p}}{x_i}
    + \phi\dpd{}{x_j}\left\{ \nu\left( \dpd{\DAvg{u_i}}{x_j}+
      \dpd{\DAvg{u_j}}{x_i} \right) \right\}
                         + \dpd{\phi\tau_{ij} }{x_j} \nonumber\\
    &+& \tilde{F}_i -\nu \mathcal{K}^{-1}\left(1 + \mathcal{F}\right)\phi^2\DAvg{u_i}.
  \label{eq:modif_mom_drag}
\end{eqnarray}

According to the analysis by \citet{breugem2005}, 
\annotatednew{for turbulent flows in porous media} the sub-filter stresses can \annotatednew{usually}
be neglected in the porous region\annotatednew{, as the contributions of drag to the momentum
balance is greater than the turbulent dispersion. 
Furthermore, in most cases, it can be assumed that the turbulence is confined to a pore size. 
This assumption, expressed as the pore scale prevalence hypothesis (PSPH),
has been tested in several DNS studies, see for example work by~{\citet{jin2015}}. 
Therefore, if the turbulence does not lead to significant mixing on the
macroscopic scale, than there is no need to add any up-scaled dispersion
model (e.g.~additional viscosity) to the equations.}
Additionally, a DNS
study \annotatednew{of the flow over a modelled porous medium by}~\citet{breugem2006} suggests that resolved fluctuations diminish quickly
in the \annotatedchange{modeled porous medium}{porous material}.
\annotatedchange{Hence}{Having that in mind}, for LES simulations, we decided to use
eddy-viscosity closure for the sub-filter stresses (based on the Boussinesq
hypothesis~\cite{sagaut2006}) in the whole domain\annotatedchange{,}{.}
\annotatedchange{assuming that undisturbed}{Based on results from
Ref.~{\onlinecite{breugem2006}} we assume that the undisturbed, non-fluctuating}
flow field in the porous region \annotatednew{implicitly} results in a null
sub-grid scale viscosity%
\annotatedchange{(in the equation below both viscosities are combined into
\(\nu_\mathrm{eff}\))}{.} 
\annotatednew{The advantage of such an approach is its locality, i.e.~the
turbulence model remains the same for each region of the flow.}
The resultant momentum equation \annotatednew{(both viscosities are combined
into {\(\nu_\mathrm{eff}\)})},
\begin{equation}
    \phi\dpd{\DAvg{u_i}}{t}+
    \dpd{\phi\DAvg{u_i}\DAvg{u_j}}{x_j} 
    =
    \phi\dpd{\DAvg{p}}{x_i}
    + \phi\dpd{}{x_j}\left\{\nu_\mathrm{eff}\left( \dpd{\DAvg{u_i}}{x_j}+
      \dpd{\DAvg{u_j}}{x_i} \right) \right\}
  + \tilde{F}_i -\nu \mathcal{K}^{-1}\left(1 + \mathcal{F}\right)\phi^2\DAvg{u_i},
\end{equation}
leads to the semi-discrete equation for velocity similar to the one obtained from
the standard incompressible Navier--Stokes equation:
\begin{equation}\label{eq::vel-disc}
  a_p \DAvg{\vec{u}}_p = \tnsr{H}(\DAvg{\vec{u}}) - \phi\nabla \DAvg{p}.
\end{equation}
When interpolated onto faces and inserted into the continuity equation, 
the standard incompressible Navier--Stokes equation forms the
Poisson equation for the pressure \citep{moukalled2016}:
\begin{equation}\label{eq::press-disc}
  \nabla\cdot \left( \frac{\phi_{VT}^2}{a_p}\nabla \DAvg{p}\right) 
  = \nabla \cdot \left( \frac{\phi_{VT}\tnsr{H}(\DAvg{\vec{u}})}{a_p} \right) - \Lambda^c.
\end{equation}
Equations~\eqref{eq::vel-disc} and~\eqref{eq::press-disc} are solved in an
identical fashion as in the standard PIMPLE-based OpenFOAM solvers. 

The algorithm for the steady state solver is constructed in a similar way, starting
again from Equation~\eqref{eq:dans_steady_por} with the time derivative in the
momentum equation set to zero. As mentioned before, instead of models, an accurate 
representations of unclosed terms, obtained from explicit filtering, is used.

\subsection{Simulations setup}

All of the simulations and postprocessing was performed using the OpenFOAM
toolbox \citep{weller1998, jasak1996}. For the PRSs, the divergence term has
been discretised with the central difference scheme. Stationary computations
were carried out with SIMPLE steady-state algorithm \citep{patankar1972,
moukalled2016}. The LES computations were performed with WALE \cite{nicoud1999}
model using the incompressible, pressure-based PIMPLE method. For the time
derivative, a second-order accurate backward difference scheme has been used. 

To judge the quality of the LES, the ratio of resolved to total
\emph{turbulent kinetic energy} (TKE) was used as an indicator, given by
\begin{equation*}
  M = \frac{\overline {u_i^\prime u_i^\prime }}
      {(2k_\mathrm{sgs} + \overline{u_i^\prime u_i^\prime})},
\end{equation*}
where the subgrid-scale contribution was estimated as
\(k_{sgs} = {\nu_t^2}/({C_k^2 \Delta^2})\), with \(C_k = 0.094\). The cell size
\(\Delta\) was computed as a cube root of the cell volume. A proper LES is
characterised by resolving more than 80\% of the energy spectrum
\citep{dimare2014, celik2005}, therefore, time average of \(M\) should be
greater than \(0.8\).

In case of the simulations performed with the DANS solver described in
\ref{sec:dans_solver}, the same schemes and models were used for LES as for
PRS. In case of steady-state simulations, we do not model the drag or
sub-filter stresses, instead, we use accurate values computed from explicit
filtering. The convective term was also discretised with second order accurate
upwind scheme to ensure stability in the vicinity of sharply rising source
terms.

\section{Test cases}\label{s:test-cases}

In the present work, we consider an incompressible flow in two geometries,
where both flow fields are simulated and analysed in a similar fashion. In each
case, the particle-resolved simulations have been performed and filtered to
generate the CEs and other terms arising in double-filtered equations. Both
cases have also been simulated assuming that the porous medium is modelled by a
porosity distribution. A summary of the cases is provided in
Table~\ref{tab:cases-summary}.

\begin{table}[th!]
  \centering
  \caption{Summary of the test cases, filtered data sets and their naming. For
    the simulations both the geometry (t.d.\ denotes a simulation conducted in
    a truncated domain shown in Figure~\ref{fig:domain} with red lines) and the
    numerical approach is listed. Additionally, the CE source terms used to
    compute the flow are reported. The results are compared to the explicitly
    filtered results datasets from both PRSs. Here employed filter size and type is
    reported along with evaluated CEs.\label{tab:cases-summary}}
  \begin{tabular}{llllp{6.3cm}}
    \toprule
    \multicolumn{5}{c}{Simulations}\\ \midrule 
    Name     & Geometry                     & Approach & CE terms 
             & Additional information\\\midrule

    PRS-G1   & Figure~\ref{fig:domain}      & PRS      
             & & 2D, laminar, steady flow around square cylinder\\

    PRS-G2   & Figure~\ref{fig:channel_fig} & PRS/LES  
             & & 3D, turbulent, channel with a porous wall \\

    DA-G1-A  & Figure~\ref{fig:domain}, t.d.    & DANS     
             & & G1 geometry, with filtered equations \\

    DA-G1-B  & Figure~\ref{fig:domain}      & DANS    
             & & tests the importance of Eq.~\eqref{eq::Fb}, \eqref{eq::Lb} \\

    DA-G1-C1 & Figure~\ref{fig:domain}, t.d.     & DANS     
             & \(\Lambda^{\ell_V}\),  \(\vec{F}^{\ell_V}\), \(\tnsr{D}^{\ell_V}\)
             & \multirow{2}{*}{tests the importance of Eq.~\eqref{eq::FparL},
               \eqref{eq::LparL} and \eqref{eq::DparL}}\\

    DA-G1-C2 & Figure~\ref{fig:domain}, t.d. & DANS     
             &  \(\Lambda^{\ell_V}\) & \\

    DA-G2-A1 & Figure~\ref{fig:channel_fig} & DANS/LES 
             &  &\multirow{2}{*}{G2 geometry, drag fully modelled, %
             \(\ell_V\) like below} \\

    DA-G2-A2 & Figure~\ref{fig:channel_fig} & DANS/LES 
             &  &  \\
    \midrule

    \multicolumn{5}{c}{Filtered data sets}\\ \midrule 
    Name  & \(\ell_V\)         & Kernel & CE terms 
          & Additional information\\\midrule

    G1-A  & \(4\sqrt{\pi/6}a\) & Eq. \eqref{eq:gaussian} 
          &  & filtering done in the truncated domain\\

    G1-B  & \(4\sqrt{\pi/6}a\) & Eq. \eqref{eq:gaussian} 
          & \(\Lambda^{b}\),  \(\vec{F}^{b}\) 
          & computation of boundary CEs\\

    G1-C  & inhomogeneous       & Eq. \eqref{eq:gaussian} 
          & \(\Lambda^{\ell_V}\),  \(\vec{F}^{\ell_V}\), \(\tnsr{D}^{\ell_V}\) 
          & computation of inhomogeneous filtering CEs\\

    G2-A1 & \(H/10\)           & Eq. \eqref{eq:celular}  
          &  & filtered as in Ref~\onlinecite{breugem2005} \\

    G2-A2 & \(H/5\)            & Eq. \eqref{eq:celular}  
          &  & tests the influence of filter size vs.\ G2-B\\

    G2-B  & inhomogeneous       & Eq. \eqref{eq:gaussian} 
          & \(\Lambda^{\ell_V}\),  \(\vec{F}^{\ell_V}\) 
          & change of filter size at porous-fluid interface \\

    \bottomrule
  \end{tabular}
\end{table}

\subsection{Two-dimensional flow over a square cylinder (G1)}

\begin{figure}[b!]
  \centering
  \begin{minipage}{.58\textwidth}
    \includegraphics[width=\linewidth]{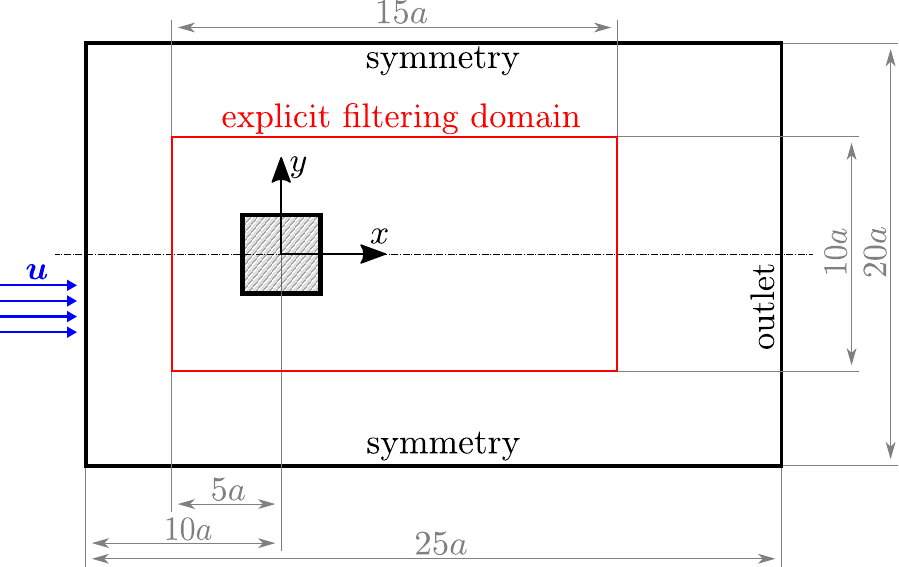}
    \caption{Schematic depiction of computational geometry and boundary
      conditions for the PRS-G1 simulation. Red lines denote the truncated
      domain, used for explicit filtering without commutation error terms and
      DA-G1-A, C1 and C2 simulations.}
    \label{fig:domain}
  \end{minipage}\hfill%
  \begin{minipage}{.4\textwidth}
    \includegraphics[width=\linewidth]{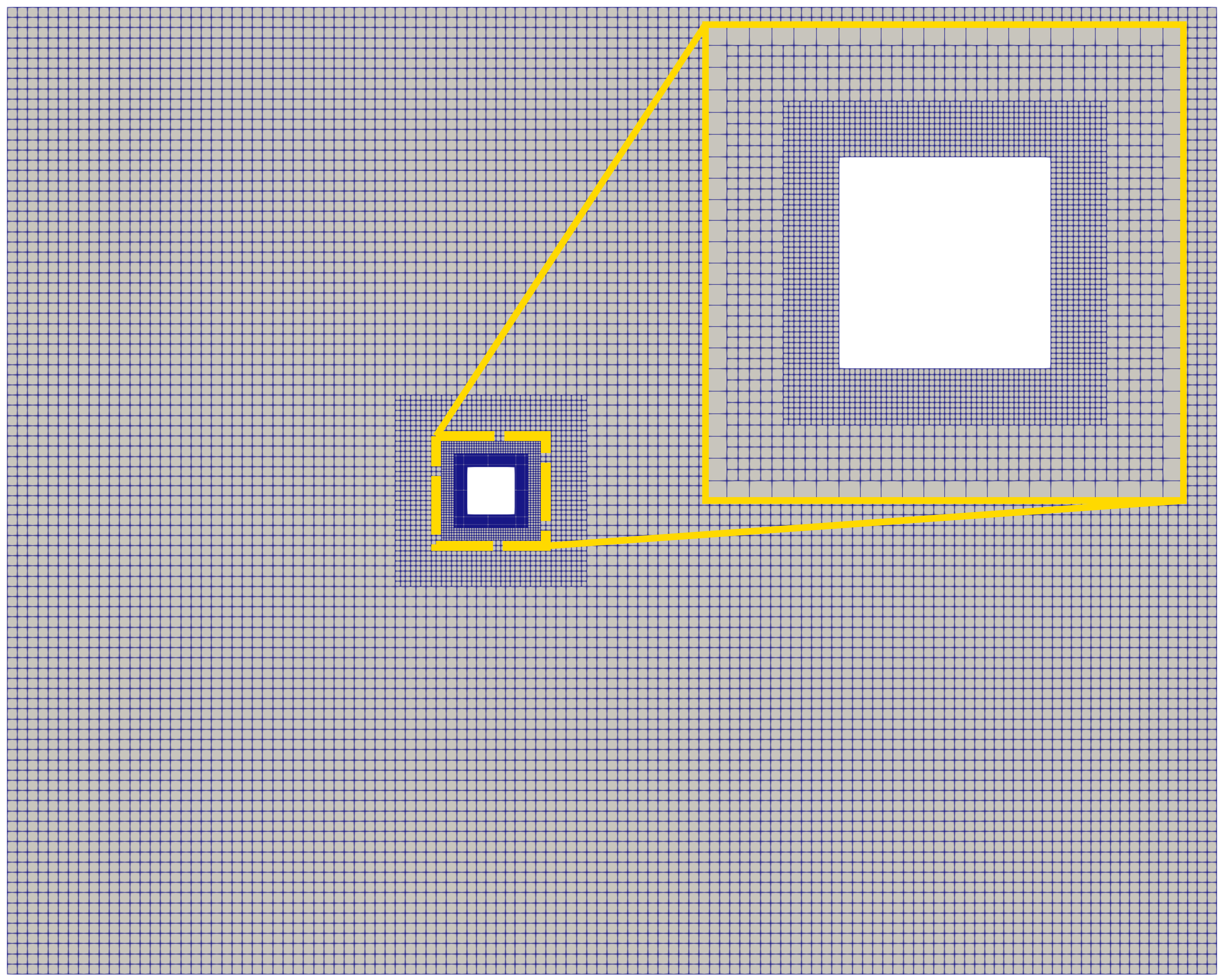}
    \caption{Computational mesh used in PRS-G1 case. Every second grid line is shown.}
    \label{fig:mesh_2d}
  \end{minipage}
\end{figure}

The first test case is a simplified PRS considering two-dimensional, laminar,
stationary flow around a square bar, denoted as PRS-G1 in this paper. The
geometrical configuration and boundary conditions are shown in Figure
\ref{fig:domain}. A uniform velocity profile is assumed at the inlet. The
Reynolds number, based on the square side \(a\) and the inlet velocity, is
equal to 1. The computational mesh used for this simulation is shown in
Figure~\ref{fig:mesh_2d}. The cell sizing is progressively refined from
\(a/10\) in the farfield region to \(a/80\) in the boundary layer of the
cylinder.

The flow has been filtered three times, resulting in three separate sets of
filtered quantities, discussed in detail in Section~\ref{s:ex_fil_G1}. First,
to establish a reference, uniform filter width was chosen resulting in a flow
field without the commutation errors. Moreover, averaging was done in a
truncated domain to avoid the influence of the boundaries. Second, the
filtering domain was expanded to introduce the boundary errors into the
equations. Third, using the truncated domain again, inhomogeneous filtering was
conducted. Those results sets are named G1-A, G1-B and G1-C respectively.

The particle-resolved simulation requires no additional modelling applied to
the Navier-Stokes system, which aids in evaluating the viscous stress and,
therefore, the drag term \(\bm{F}\) exactly. Two-dimensionality simplifies the
potentially computationally intensive non-local filtering operation. Lastly,
since the definition of the filtering is robust, without a loss of generality
we can assume that the width of the time kernel \(H\) is infinite. Since
consecutive space-time filtering is equivalent to double-filtering and a
stationary solution can be seen as time-filtered over an infinite averaging
window, we can restrict our investigation to spatial filtering of the
stationary solution. Therefore, while conducting inhomogenous filtering, we
assume that the terms \(\vec{F}^{T_0}\) and \annotatedchange{\(C^{T_0}\)}{\(\Lambda^{T_0}\)} are equal to zero. 

This flow has also been computed employing the solver described in
Section~\ref{s:numerics}. Several variants of the simulation have been
performed, each focused on testing the importance of different error terms
while implementing the equations. In these simulations, the CEs are not being
modelled, instead, the fields from results sets A to C are used. The case
DA-G1-A aims to reproduce velocity and pressure fields from G1-A and is
computed in the same truncated domain. Simulation DA-G1-B was conducted in the
full domain and investigates the importance of \(\Lambda^b\) and \(\vec{F}^b\)
terms. Similarly, DA-G1-C1 and C2 test the commutation error terms related to
inhomogeneous filtering. The particularities of each setup are described in
detail in Section~\ref{s:results}.

The simulations of DA-G1 test case were conducted on the purely quadrilateral
uniform meshes. Cases DA-G1-A1, B1 and B2 had cell size set as \(0.08a\) in the
horizontal direction and \(0.1a\) in the vertical direction. The cases C1 and
C2, following the results of the sensitivity study described in
Appendix~\ref{s:sens_study}, were meshed with cell size set to \(0.05a\).

\subsection{Turbulent flow in a channel with permeable wall (G2)}
The second case considered in the study, is the turbulent flow in an infinitely
long planar channel, with the lower section of the geometry occupied by a set
of cubes, thereby modelling a sparse packing or porous medium. This case, named
PRS-G2, is based on the work of \citet{breugem2005} who conducted both a
particle-resolved and homogenised DNS of this flow. As mentioned in the
introduction, if such a configuration is simulated with a porosity-based drag
model, it has to be ``blended'' with the standard Naver-Stokes equations. This
in turn requires the filter size to change and induces the commutation error.
While performing the DNS with a drag model, \citet{breugem2005} neglected the
commutation error. We use his results as a reference, adapt his modelling
approach to work in the LES framework and evaluate the commutation error in a
more realistic setting: unsteady, turbulent flow. 

\begin{figure}[b!]
  \centering
  \begin{minipage}{.47\textwidth}
    \centering
    \includegraphics[width=\textwidth]{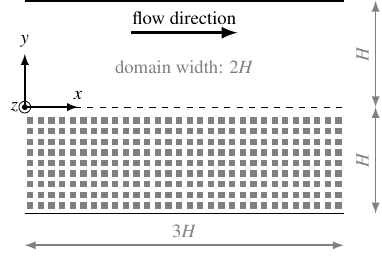}
    \caption{Schematic depiction of computational geometry for the 
    turbulent flow in the channel with the lower part occupied with porous matrix 
    modelled by a set of cubes.}
    \label{fig:channel_fig}
  \end{minipage}\hfill
  \begin{minipage}{.47\textwidth}
    \centering
    \includegraphics[width=0.8\textwidth]{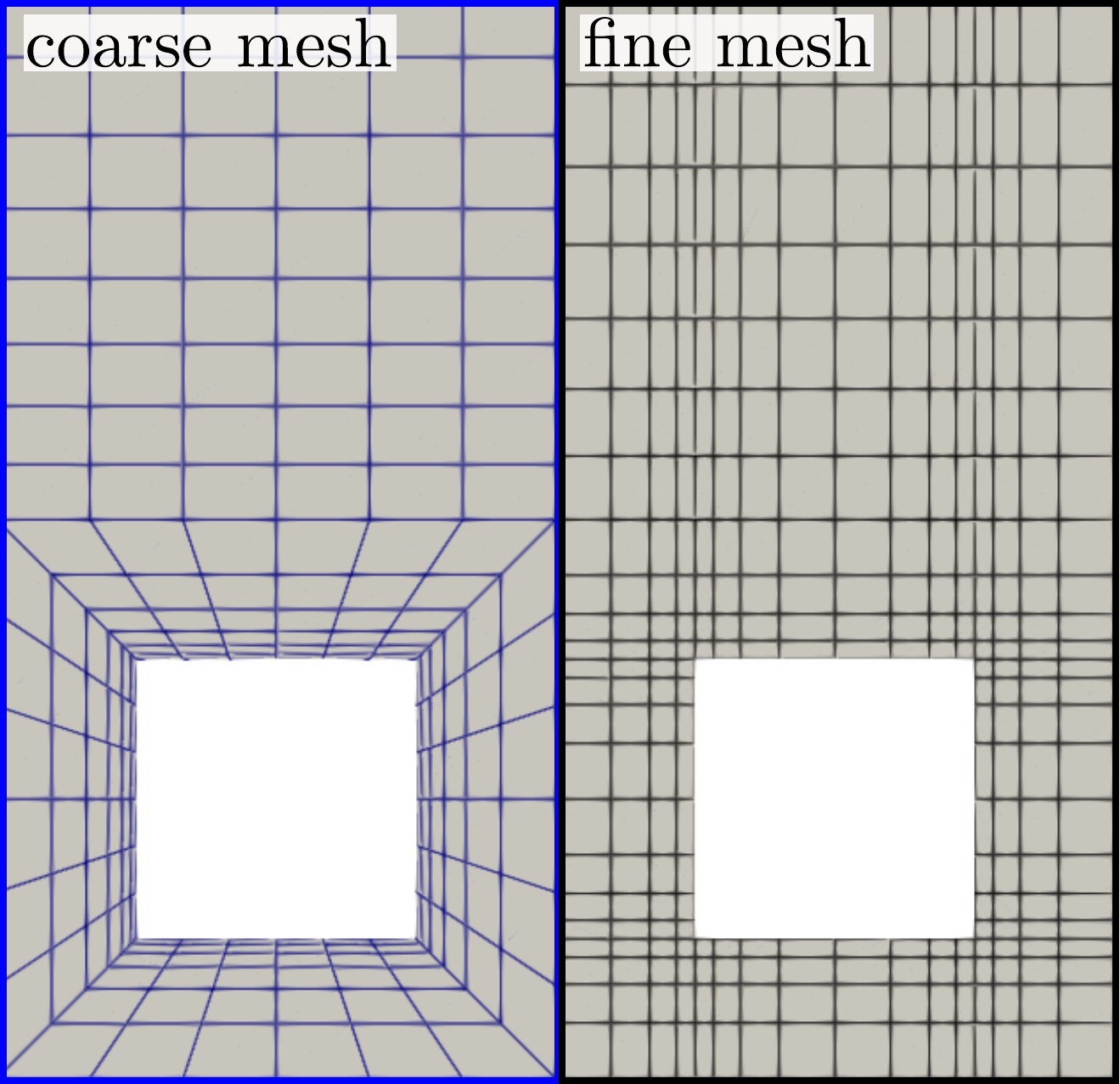}
    \caption{Detailed view of the computational meshes used for the
      channel simulation near the last layer of cubes (slice through the middle
      of the cube). Only one column of cubes is shown for each mesh.}
    \label{fig:meshes}
  \end{minipage}\hfill
\end{figure}

The geometry of the computational domain is visible in Figure
\ref{fig:channel_fig}. The Reynolds number is defined as \(Re_\mathrm{bulk} =
U_\mathrm{bulk}H/\nu= 5500\) (bulk velocity is computed in the channel section
\(y\geq0\)). The coordinate system is defined in such a way that \(x\) is the
streamwise dimension, \(y\) is wall-normal, and \(z\) is spanwise. The channel
has dimensions of \(3H\times2H\times2H\). The domain is periodic in both
streamwise and spanwise directions. The cubes have the size of \(H/20\). The
distance between the centres of the cubes is equal to \(H/10\). The first cube's
centre is located \(3H/40\) from the bottom wall. There are 9 layers of
cubes, leaving the centre of the top layer at \(y = -0.125H\). Chosen
dimensions of the geometry give the nominal porosity in the bulk \(\phi =
0.875\).
The flow is driven by a momentum source, prescribing the
pressure gradient so that the bulk velocity in the upper part of the channel
results in the prescribed \(Re_\mathrm{bulk}\).

The unsteady fields were time-averaged over the period required for 60
flow-trough times. To further improve the convergence of statistics, the time
averaged fields were spatially averaged considering the periodic nature of the
computational domain. The geometry contains 600 identical sections, repeated
in \(x\) and \(z\) directions, consisting of a column of cubes and a channel
section on top. 

The computations on two different meshes have been realised to ensure a proper
resolution of both the turbulent flow field and the stress tensor at the walls
of the cubes. The latter is of great importance, as the values at the surfaces
will contribute directly to the computation of the drag force using explicit
filtering. Both meshes, the coarse and the fine one, are visible in
Figure~\ref{fig:meshes}. The statistics of the meshes are also presented in 
Table~\ref{tab:mesh_stats}.

According to \citet{chapman1979}, to obtain an appropriate representation of
structures inside the boundary layer, the wall spacing needs to fulfil the
requirements \(x^+ < 100\), \(y^+ < 2\), and \(z^+ < 20\). Both of the
considered meshes, fulfil these criteria. 

\begin{table}
  \centering
  \caption{Statistics of the meshes used to conduct the particle-resolved LES
    of configuration G2. Cell sizes \(\Delta_x^{+t}\), \(\Delta_y^{+t}\) and
    \(\Delta_z^{+t}\) are normalised using the friction velocity computed at
    the top wall of the channel, \(u_\tau^t\).}
  \label{tab:mesh_stats}
  \begin{tabular}{lcc}
    \toprule
    mesh & coarse & fine \\\midrule
    number of cells & \num{8467200} & \num{29952000} \\
    max non-orhogonality angle & \SI{50}{\degree} & \SI{0}{\degree} \\
    max skewness & \num{2.5} & \num{0} \\
    \(\max x^+, y^+, z^+\) & \num{8.6}, \num{0.7}, \num{8.6} & \num{7.8}, \num{1.4}, \num{6.1}\\
    clear channel \(\Delta_x^{+t}\), \(\Delta_y^{+t}\), \(\Delta_z^{+t}\) 
      & \num{6.5}, \num{0.65}-\num{15.3}, \num{6.5} 
      &  \num{1.3}-\num{4}, \num{0.55}-\num{7.67}, \num{1.3}-\num{4} \\\bottomrule
  \end{tabular}
\end{table}

The resolved fields have been filtered with both Cellular and Gaussian kernels,
Equations \eqref{eq:celular} and \eqref{eq:gaussian} respectively. Results from
filtering with the first are used to validate the simulation against
Ref~\onlinecite{breugem2005}. The latter is employed to investigate the CE, due
to the smoothness of the \(\partial \widehat{G} / \partial\ell_V\) function,
given by Equation~\eqref{eq:gaussian_derivative}. Since the time-filtering
kernel assumes an infinite time scale, only the commutation errors related to
spatial filtering were considered.

\begin{figure}[b]
  \centering
  \includegraphics[width=0.45\textwidth]{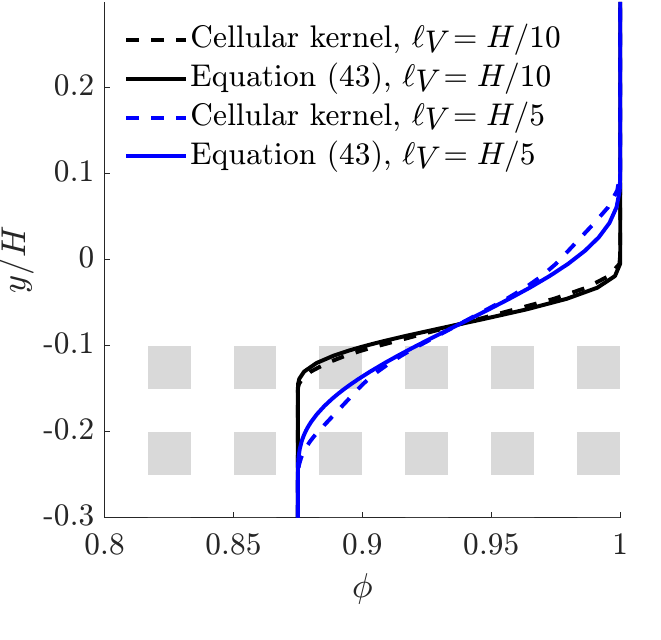}
  \caption{Distributions of porosity in DA-G2-A and B simulations (for two
  different filter sizes), compared to curves computed by explicit filtering
  with Cellular kernel.}
  \label{fig:porosities}
\end{figure}

The flow in a channel was also computed with the implemented solver. The
simulations (named DA-G2-A and B) were performed assuming two different filter sizes
inside the packing, \(\ell_V = H/10\) and \(\ell_V = H/5\), resulting in two
different porosity distributions in the channel \(\phi(y)\) given by:
\begin{equation}\label{eq:porosity_distr}
  y \in (\delta^{-}, \delta^{+}):\quad
  \frac{\phi(y)}{\phi^+ - \phi^-} 
  = 6{\left(\dfrac{y - \delta^+}{\delta^+ - \delta^-}\right)}^5
  + 15{\left(\dfrac{y - \delta^+}{\delta^+ - \delta^-}\right)}^4
  + 10{\left(\dfrac{y - \delta^+}{\delta^+ - \delta^-}\right)}^3
  + \frac{\phi^+}{\phi^+ - \phi^-}.
\end{equation}
Porosity over the bed, \(y\geq\delta^+\), denoted as \(\phi^+\) is equal to 1.
In the porous region, for \(y\leq\delta^-\), it is given as \(\phi^- = 0.875\).
For \(\ell_V = H/10\) the porosity changes in the region \((\delta^-, \delta^+)
= (0.075H,0)\), whereas in case of \(\ell_V = H/5\) it changes over \((-0.25H,
0.1H)\). The profiles of the porosities are depicted in Figure
\ref{fig:porosities}. They provide a very good approximation of \(\phi\)
resulting from the filtering procedure.

The dimensions of the domain remain the same as in the particle-resolved
simulation. The flow is also driven by a pressure gradient updated to ensure
the correct \(Re_\mathrm{bulk}\). The drag closure used for these configurations and
equations used in both simulations is described in Section~\ref{s:numerics}.

Because the cases DA-G2-A and B employ LES as the turbulence modelling strategy,
we assume that the time filtering kernel is reduced to the \(\delta\) function.
We can further assume that the spatial filter is connected to the mesh size in
the clear fluid region, \(\phi=1\). On the other hand, in the porous region, we
only assume spatial filtering with the \(\ell_V=H/10\) or \(\ell_V=H/5\). 

To reach the
double-averaged description of flow parameters, the solutions have been
additionally time averaged across 60 flow-through times. To further boost
collected statistics, the flow field can be averaged in homogeneous
directions. 

The simulations have been
performed on fully hexahedral and orthogonal meshes, refined near the walls
and interface region. For each case, two mesh resolutions were tested to assess
the influence of mesh resolution on the results. The statistics of the meshes
are listed in Table \ref{tab:mesh_stats_va}. LES quality was also tested with
resolved energy criterion.

\begin{table}
  \centering
  \caption{Statistics of the meshes used for simulation of the G2 case with the
    double-averaged equations. Cell sizes are normalised by the friction
    velocity at the top wall of the channel \(u_\tau^t\). The last column describes
    the wall-normal cell dimension in the area where porosity changes.}
  \label{tab:mesh_stats_va}
  \begin{tabular}{lccc}
    \toprule
    mesh \& \(\ell_V\) & number of cells 
          & \(\Delta_x^{+t}\), \(\Delta_y^{+t}\), \(\Delta_z^{+t}\) & interface \(\Delta_y^{t+}\) \\\midrule
    coarse \& \(H/10\) & \num{600000} & \num{8}, \num{0.4}-\num{13.6}, \num{15.5} & \num{3.5}\\
    coarse \& \(H/5\) & \num{600000} & \num{8}, \num{0.4}-\num{14}, \num{16} & \num{3.6}\\
    fine \& \(H/10\) & \num{2700000} & \num{8}, \num{0.2}-\num{6}, \num{8} & \num{2}\\
    fine \& \(H/5\) & \num{2400000} & \num{8}, \num{0.3}-\num{9}, \num{8} & \num{2.7}\\
    \bottomrule
  \end{tabular}
\end{table}

\section{Results}\label{s:results}
\subsection{Explicit filtering: G1 geometry}\label{s:ex_fil_G1}

\begin{figure*}[b]
  \centering
  \includegraphics[width=\linewidth]{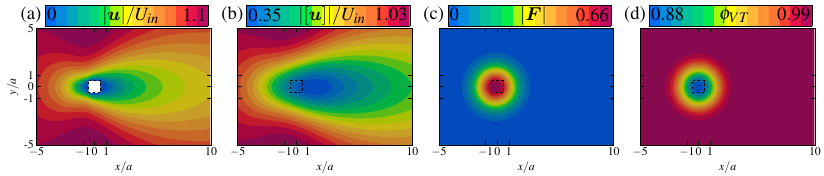}
  \caption{Distributions of various terms computed from case G1-A, (a)
  unfiltered velocity field, (b) filtered velocity field,
  (c) magnitude of computed drag force \(\vec{F}\), (d) porosity distribution.}
  \label{fig:small_filtered_4}
  \subfloat{\label{fig:vel_resolved}}
  \subfloat{\label{fig:small_vel}}
  \subfloat{\label{fig:small_F}}
  \subfloat{\label{fig:small_por}}
\end{figure*}

The PRS has been filtered to obtain \(\vec{u}\), \(p\), the product
\(\vec{u}\vec{u}\) and \(\tnsr{\tau}\). Additionally, the drag term \(\vec{F}\)
was evaluated using the pressure and derivatives of velocity on the boundaries.
At first, these operations were done on a smaller filtering domain (depicted in
Figure~\ref{fig:domain} with red lines), to separate the errors related to explicit
filtering and the computation of the unclosed terms from the boundary-related commutation
errors. The computed fields constitute the results set G1-A. The filter size was
set as \(\ell_V=4\sqrt{\pi/6}a\). A comparison of filtered and reference
velocity and pressure along the \(x\)-axis can be seen in
Figure~\ref{fig:fields_small}. Additionally, the distribution of resolved and
filtered velocity fields is visible respectively in
Figures~\ref{fig:vel_resolved} and~\ref{fig:small_vel}. The distribution of
filtered velocity is visibly smeared, however, both filtered fields approach
their unfiltered values far away from the obstacle. 

The error related to clipping of the filtering molecule manifests itself as
inaccurately computed porosity (Figure~\ref{fig:small_por}), which never
reaches a value of 1 in the free fluid. The magnitude of \(\vec{F}\) is presented in
Figure~\ref{fig:small_F}. To test the accuracy of the explicit filtering
procedure, the residual of the double-averaged momentum equation was computed. Since no
filtering near boundaries occurs and \(\ell_V = \text{const.}\) all of the
commutation error terms are set to 0. The error of the momentum equation is
normalised with the maximum imbalance of the stationary equations without the
source term, i.e.\ \(\max |(\partial \DAvg{u_i u_j} / \partial x_j + \partial
\DAvg{\sigma_{ij}} / \partial x_j)|\), and denoted \(\vec{e}_{\text{mom}}\). 
Its maximal and average values are presented in Table~\ref{tab:error_table}.
Since there is no need to evaluate source terms for the continuity equation for
this filtering domain, normalised values of continuity errors were not
computed. 
\begin{figure*}[b]
  \centering
  \includegraphics[width=\linewidth]{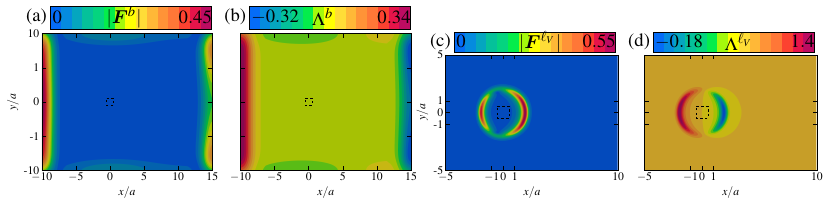}
  \caption{Distributions of commutation error terms computed from cases G1-B and
  G1-C, (a) magnitude of boundary error in momentum equation, (b) boundary error
  in continuum equation (c) magnitude of inhomogeneous spatial filtering error in
  momentum , (d) inhomogeneous spatial filtering error in continuum equation.}
  \subfloat{\label{fig:fig_Fb}}
  \subfloat{\label{fig:big_Cb}}
  \subfloat{\label{fig:vd_Fgrad}}
  \subfloat{\label{fig:vd_Cgrad}}
\end{figure*}


\begin{table*}[th!]
  \centering
\caption{Errors resulting from the  evaluation of momentum and continuity
  residuals in all the three cases: G1-A - filtering without commutation
  errors on the clipped filtering domain, G1-B - filtering with boundary
  commutation errors, G1-C - inhomogeneous filtering on the small
  domain.\label{tab:error_table}}
\begin{tabular}{lcccc}
  \toprule
  Case& \(\max \; e_{\text{mom}}\) & \(\max \; e_{\text{con}}\) 
      & \(\text{avg}\; e_{\text{mom}}\) & \(\text{avg}\;e_{\text{con}}\)\\
  \midrule
  G1-A & 0.22\%  & -       & 0.019\%   & -\\
  G1-B & 0.23\%  & 0.32\%  & 0.004\%   & 0.034\%\\
  G1-C & 0.97\%  & 0.86\%  & 0.025\%   & 0.05 \% \\
  \bottomrule
\end{tabular}
\end{table*}
Both errors related to filtering near the boundary are mathematically the same
as the drag force term, so for results set G1-B, they can be evaluated in the same
fashion as \(\vec{F}\). To capture those errors, filtering has to be conducted
in the same domain in which the reference simulation took place. Both
\(\vec{F}^b\) and \(\Lambda^b\) are connected to a decrease of porosity near the
edges of the domain (porosity decreases because only part of the kernel
overlaps the computational domain). Since the divergence of the product
\(\phi\DAvg{\vec{u}}\) is the main term in the continuity equation, a
decrease in porosity should lead to a forced change of velocity near the
boundary. However, since the intrinsic velocity is weighted by the porosity
field, it remains similar to the unfiltered distribution, and \(\Lambda^b\) field is
required to satisfy the continuity balance.

Distributions of the computed terms are shown in Figures~\ref{fig:fig_Fb} and
\ref{fig:big_Cb}. The largest values of the sources are located near the inlet
and outlet of the domain, consistently with the intuitive interpretation
provided above. The residual of the DANS equations is also evaluated in this
study and presented in Table~\ref{tab:error_table}. The \(e_{\text{con}}\)
value is the absolute imbalance of the continuity equation normalised by the
maximum of \(|\partial \phi\DAvg{u_i}/\partial x_i|\). The errors are
comparable to the previous case, signifying a proper resolution of both source
terms.


These results also substantiate our observations, related to the possible treatment
of non-solid boundaries of the domain in the solver implementing DANS
equations, described in Section~\ref{s:comm_err}. Since the velocity can be
prescribed properly at the inlet (the same boundary condition could be used as for
the unfiltered field), there is no reason to assign values of porosity smaller
than 1 near non-solid or non-permeable boundaries. Therefore, this decrease in
the computed value of porosity can be deemed artificial and such a solver should
not require \(F^b\) and \(C^b\) terms to reproduce explicitly filtered results.

Lastly, the terms \(\vec{F}^{\ell_V}\) and \(C^{\ell_V}\) were evaluated.
The computed set of filtered fields was named G1-C. To isolate them from the
influence of boundary-related errors, a smaller domain was used for the
filtering. The distribution of \(\ell_V\) was given by the expression \(\ell_V
=\sqrt{\pi/6}( -3a\arctan(4\sqrt{x_i x_i}/a - 8 )/\pi + 5/2 a ) \). It varies
smoothly and reaches a comparable value as in previous tests in the vicinity of
the cylinder.

Distributions of both error terms are presented in Figures~\ref{fig:vd_Fgrad}
and~\ref{fig:vd_Cgrad}. As shown in Appendix~\ref{s:sens_study} an increased
resolution was necessary to ensure proper representation of sharp gradients in the
filtered fields. Despite the increase in resolution, the errors (summarised in
Table~\ref{tab:error_table}) are still higher than in the other cases, with the
maximum relative momentum error being approximately 1\%. Nevertheless, the
momentum and continuity errors are small enough to show that all terms were
computed correctly in all three cases, thus verifying the derived expressions
for commutation errors.

\subsection{Solving DANS equations: case G1}
\begin{figure}[t]
  \centering
  \begin{minipage}{0.47\textwidth}
    \centering
    \includegraphics[width=\textwidth]{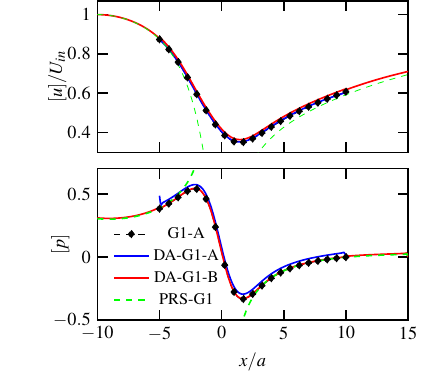}
    \caption{Plots of the streamwise velocity and pressure: from the
      particle-resolved simulation, explicitly filtered with a homogeneous
      filter and results from DANS equations.
    }
    \label{fig:fields_small}
  \end{minipage}\hfill%
  \begin{minipage}{0.47\textwidth}
    \centering
    \includegraphics[width=\textwidth]{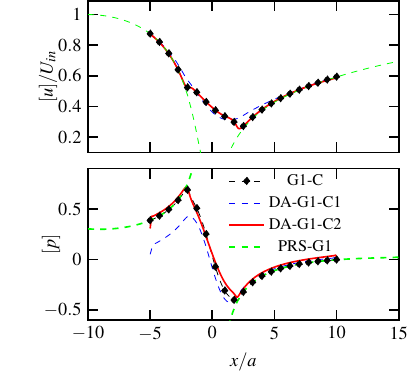}
    \caption{Plots of the streamwise velocity and pressure: from the
      particle-resolved simulation, explicitly filtered with an inhomogeneous
      filter and results from DANS equations assuming accurate or null commutation 
      error in momentum equation.
    }
    \label{fig:fields_vdelta}
  \end{minipage}
\end{figure}

The presented version of the discretised DANS equations was first
applied to simulate the flow field from case DA-G1-A. Distributions of the
explicitly filtered velocity were used as the boundary conditions in the
simulation, for the inlet, upper and lower edges of the domain. For the outlet, a
zero gradient boundary condition was employed for velocity along with a
reference value for pressure. Residuals of the equations were driven below
\(10^{-6}\) to ensure the proper convergence of the nonlinear system. Both the
\(\DAvg{\vec{u}}\) and \(\DAvg{p}\), are in very good agreement with the
explicitly filtered data. Plots of both fields along the \(x\)-axis are
presented in Figure~\ref{fig:fields_small}. The computed variables are
similar to the original fields in areas sufficiently far from the porous region
and where both \(\vec{u}\) and \(p\) are smooth. 

In case DA-G1-B1, \(\vec{F}\), \(\phi\) and \(\tnsr{\tau}\) fields from
G1-A were extrapolated and used for the computation in a domain of the same
size as used in the reference simulation. The uniform boundary condition at the
inlet and the symmetry conditions at the sides of the domain were prescribed
for the filtered velocity. Obtained solution (presented in Figure
\ref{fig:fields_small}) was in excellent agreement with filtered values. These
results support the hypothesis that boundary-related commutation errors
need not be included in the computation when the correct porosity is prescribed
near the edges of the domain.


Finally, commutation errors connected to the change of \(\ell_V\) were
implemented in the solver and tested in case DA-G1-C1. Similarly to the first test,
clipped domain and non-uniform Dirichlet conditions were used for velocity.
Plots of \(\DAvg{u}\) and \(\DAvg{p}\) are visible in Figure
\ref{fig:fields_vdelta}. The velocity field is in close agreement with the
filtered results. The main feature of the flowfield is the sharp changes of
velocity in areas where \(\ell_V\) changes are reproduced very well by the
implemented solver. The distribution of intrinsic pressure matches the
reference results as well. Additionally, values of \(\DAvg{p}\) near the outlet
and inlet are in good agreement with the reference data, aiding in evaluating
the pressure drop, one of the important quantities of interest when conducting
simulations of packed beds or porous media.

Additionally, the simulation without the commutation errors in the momentum
equation was performed. In this case, labelled DA-G1-C2, the filtered velocity
fails to account for the changes introduced by the inhomogeneous filtering.
Pressure is also reproduced inaccurately. This suggests that including
commutation error terms is necessary for the accurate computation of  inhomogeneously 
filtered results in such a simplified configuration. That said, G1-C was
computed assuming an arbitrary, fast-changing distribution of \(\ell_V\).
Examination of more realistic flow conditions is necessary to derive more
generalisable conclusions.



\begin{figure}[t]
  \centering
  \includegraphics[width=0.8\textwidth]{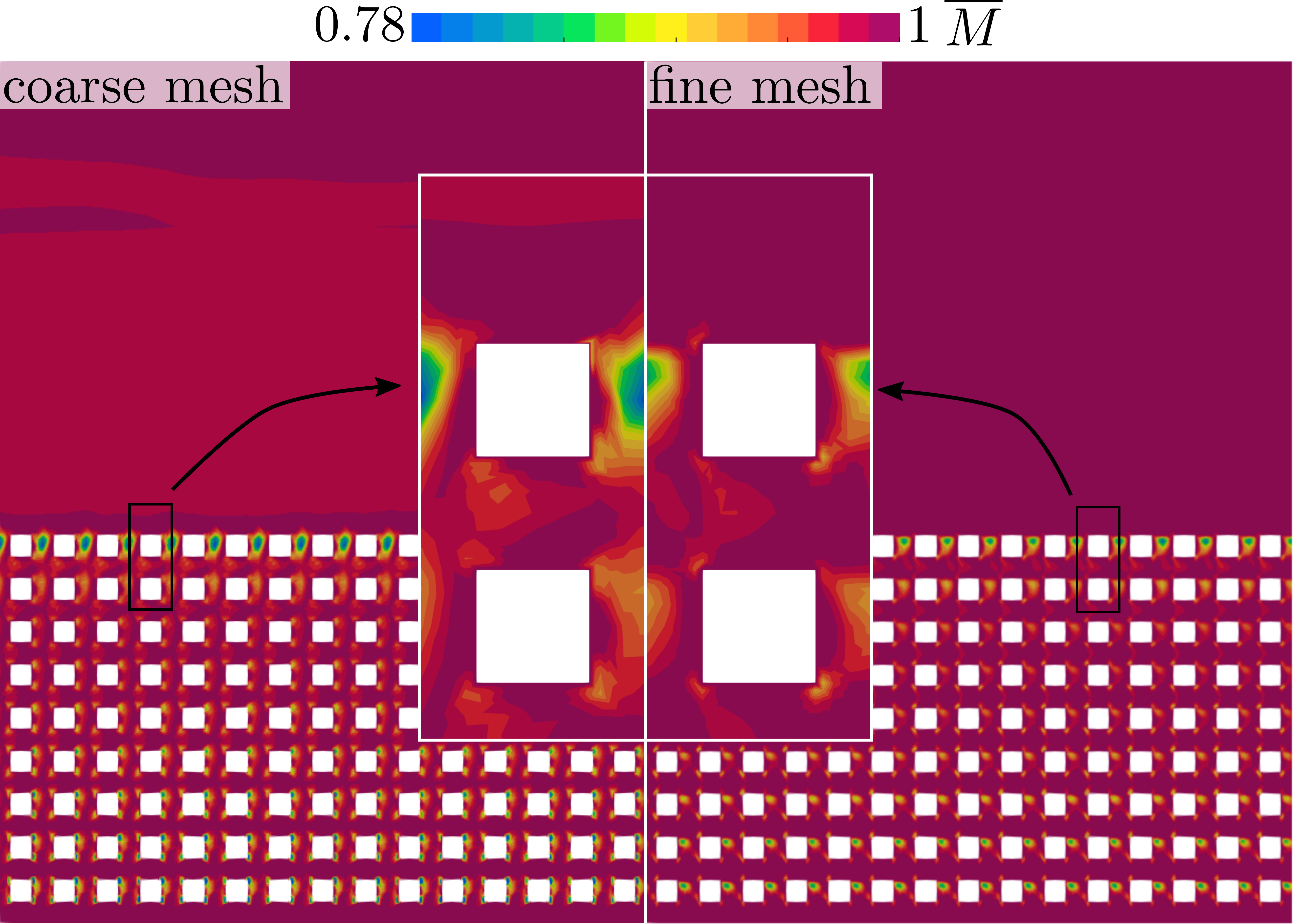}
  \caption{Time-averaged values of the LES quality criterion, based on resolved turbulent kinetic
  energy, using the coarse mesh (left) and the fine mesh (right).}
  \label{fig:M}
\end{figure}

\subsection{Validation of PRS-G2 case}

The quality of the performed LES was estimated using the resolved turbulent
kinetic energy criterion. The plot of \(\overline{M}\), including a zoomed view
at the region near the top layer of cubes, is visible in Figure~\ref{fig:M}.
The computation on the coarse mesh indicated that the chosen resolution is
insufficient, as \SI{80}{\percent} of the energy was not resolved near the
topmost layer of cubes and in the deeper parts of the porous wall. The finer
mesh ensures that the criterion of \(\overline{M}>0.8\) is fulfilled in the
whole domain. Henceforth, if it is not specified otherwise, we will limit the
discussion to the results computed on the finer mesh.

\begin{figure}[t]
    \centering
    \includegraphics[width=\linewidth]{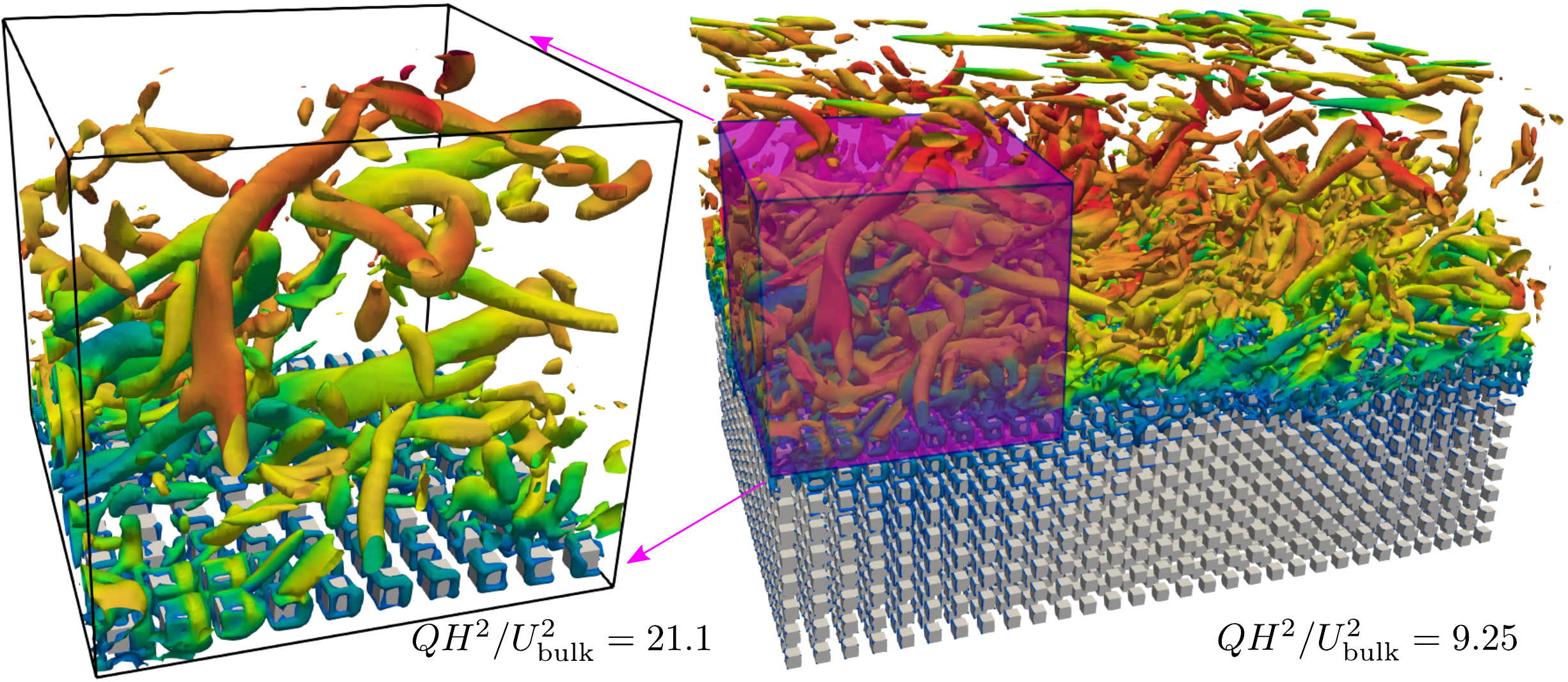}
    \caption{Q criterion iso-surface colored by the magnitude of velocity.}
    \label{fig:Q}
\end{figure}

The presence of the porous layer at the bottom influences drastically the
nature of the flow and turbulence above the cubes. Additionally, the velocity
in the porous layer is non-uniform, with the regions of faster fluid
penetrating the permeable structure periodically. Such flow is dominated by
large-scale, two-dimensional K-H instability\cite{breugem2005, kuwata2016},
which is probably responsible for the presence of the regions of the faster flow.
Figure~\ref{fig:Q} presents the visualisation of vortical structures with
\(Q\) criterion contours. The vortices spanning from the porous layer are
present in a greater number than typical structures in the boundary layer near
the top wall. Additionally, contrary to the elongated structures present in
the boundary layer near a smooth boundary, they are not aligned with the streamwise
direction, similar to the results reported by \citet{kuwata2016}.

The results have been filtered with the Cellular kernel assuming
\(\ell_V=H/10\) (forming result set G2-A1), to facilitate validation and the comparison of our results
with the work of \citet{breugem2005} who used the same filtering molecule. The
filter size was constant in the whole domain, apart from the regions near the
top and bottom of the domain. There, the kernel was progressively
shrunk to ensure that the whole filtering molecule fits in the domain.
Following \citet{breugem2005} we neglect the resulting commutation error. The
accuracy of explicit filtering was once more tested by evaluating the
one-dimensional budget of double-averaged momentum. The mean absolute value of imbalance,
computed in the region \(-0.8\geq y/H \geq 0.8\) (where
\(\ell_V=\mathrm{const.}\)) was equal to \(\num{1.4e-4}H/U^2_\mathrm{bulk}\) or
around \(0.6\%\) of the mean pressure gradient driving the flow.

The Reynolds number computed with the top wall friction velocity \(Re_\tau^t
=u_\tau^t H/\nu=\num{393.7}\) does not differ greatly from the reported value
of \num{394}. The simulation on a coarser mesh resulted in a lower value of
\(Re_\tau^t = \num{388.2}\), even though the resolution near the top wall was
sufficient for an accurate reproduction of the boundary layer flow.

\begin{figure}[t!]
  \begin{center}
    \includegraphics[width=0.95\textwidth]{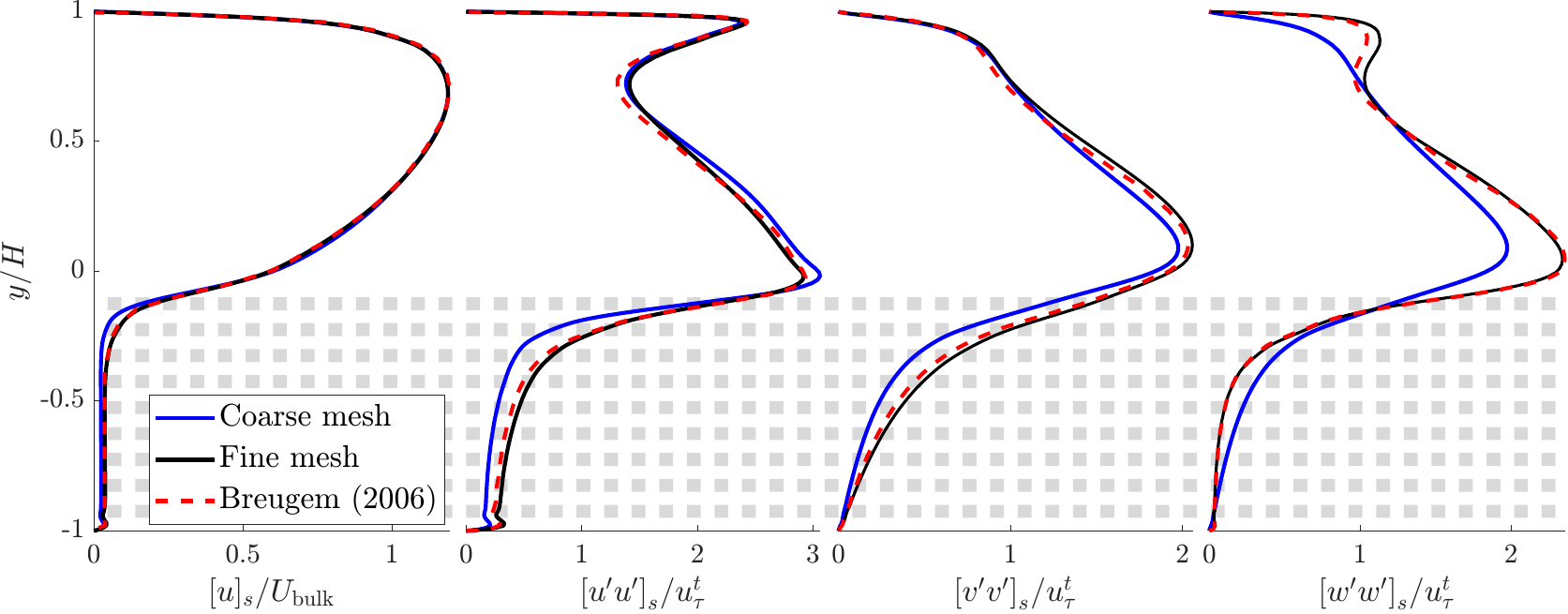}
    \caption{Filtered results from PRS-G2 case computed on the coarse and fine mesh
    (\(\ell_V=H/10\)), compared to the reference DNS\cite{breugem2005}.}
  \label{fig:channel_means_prs}
  \end{center}
\end{figure}

The comparison of double-filtered velocity profiles and velocity fluctuations
with the reference results is visible in Figure~\ref{fig:channel_means_prs}.
The velocity in the unobstructed part of the channel is predicted correctly on
both meshes, while the coarser grid underpredicts the velocity in the porous
part of the domain. Additionally, on the coarser mesh, the velocity reaches the
uniform value closer to the topmost layer of cubes.
The streamwise and wall-normal fluctuations are predicted accurately on the
fine mesh and underestimated in the porous region on the coarser mesh.


\begin{figure}[t]
    \centering
    \includegraphics[width=0.5\linewidth]{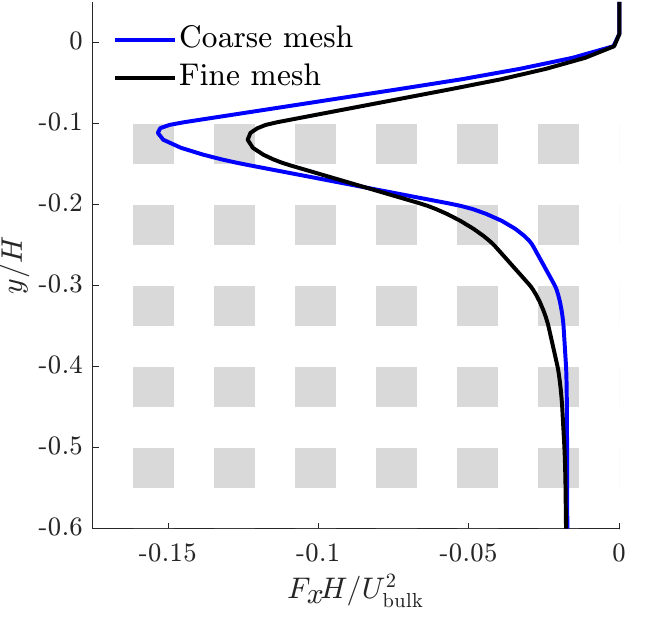}
    \caption{The drag force experienced by the fluid on the two meshes used in the 
      PRS-G2 simulation.}
    \label{fig:drag_meshes}
\end{figure}

\begin{figure}[h!]
  \begin{center}
    \includegraphics[width=0.95\textwidth]{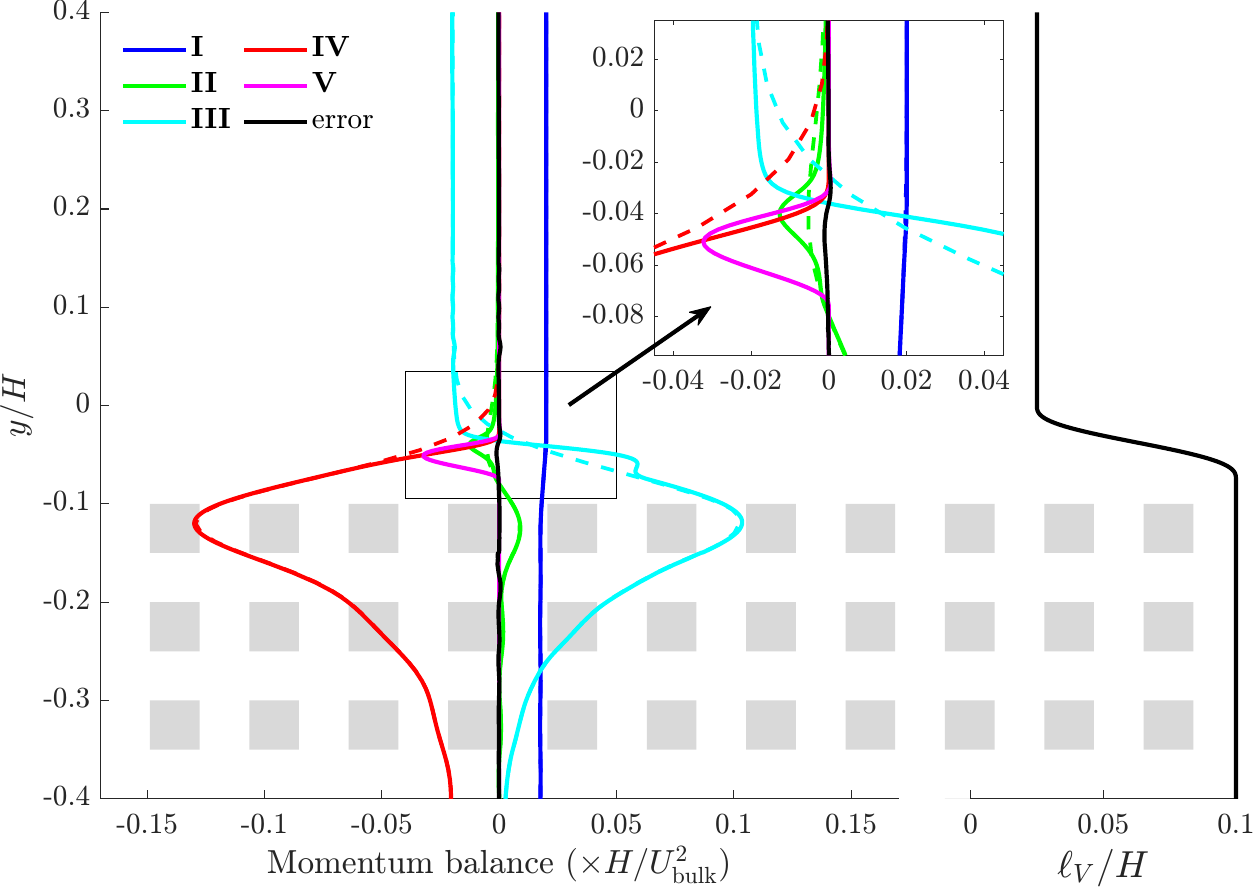}
    \caption{\textbf{Left:} space-time filtered momentum balance in the middle of the
    channel. Solid lines describe the terms defined in Eq.
    \eqref{eq:mom_balance}, evaluated with Gaussian kernel assuming changing
    filtering width. Their counterparts evaluated with the same kernel assuming
    \(\ell_V=H/10\) are displayed with dashed lines. \textbf{Right:} distribution of 
    filter size.}
  \label{fig:bal_comm_err}
  \end{center}
\end{figure}

Contrary
to the rest of the profiles, the distributions of the spanwise fluctuations
differ greatly between the two meshes. The fine mesh reproduces the DNS
results well. The coarse LES underestimates the fluctuations in both boundary
layers, while overpredicting spanwise motions inside the porous layer. 
Worse results on the coarse mesh can naturally be attributed to greater cell
spacing. However, on both meshes, the cell sizes were identical in \(x\) and
\(z\) directions, which intuitively should result in a similar accuracy of the
prediction of the streamwise and spanwise velocity fluctuations.

An alternative explanation for the inaccurately computed spanwise
fluctuations on the coarse mesh might relate to the presence of the K-H
instability. When the resolution near the top layer of cubes is
insufficient, the transfer of energy from the large-scale vortices in the
free flow to the fluid inside the porous region will not be reproduced
accurately. The generation of the spanwise motions in the boundary layers will
depend on the breakdown of wall-normal and streamwise fluctuations and is also
connected to the accuracy with which the two-dimensional K-H instability is
resolved. Therefore, insufficient resolution in the porous layer will lead to
a less accurate prediction of spanwise fluctuation in the clear channel.

Nevertheless, the fine mesh allows for sufficiently accurate LES simulation
resolving both first- and second-order statistics with good accuracy. The last
important parameter that needs to be compared between meshes is the predicted
drag force distribution. Drag forces computed from explicit filtering are
visible in Figure~\ref{fig:drag_meshes}. Since the simulation on the fine mesh
results in a correct value of \(Re_\tau^t\) and properly predicted mean
profiles, we will consider the distribution of the drag force from this
simulation as an accurate reference. In comparison, coarser mesh overpredicts
the drag force near the topmost layer of cubes which is a direct result of
increased numerical dissipation resulting from not sufficient resolution. It
provides evidence, that a slight change in resolution and gridding methodology
can influence greatly the drag force distribution and by extent the mean
velocity in the porous region. Therefore, the process of generating a grid for
a particle-resolved LES should be done with great care.

\subsection{Commutation errors at the porous-fluid interface}\label{s:comm_G2}

The particle-resolved simulation has been also used to evaluate the commutation
errors in the momentum equation. For this part of the study Gaussian function
was used as a kernel instead of Cellular weighting function. As pointed out by
\cite{quintard1994, breugem2005} and our previous work \cite{sadowski2023},
Cellular kernel is better suited to such geometry at a chosen filter size of
\(\ell_V=H/10\). However, computation of commutation errors requires the
evaluation of the kernel derivative w.r.t.\ \(\ell_V\), which would be
discontinuous in the case of Cellular kernel and introduce additional errors into
the investigation. Additionally, owing to the scaling ensured by
Eq.~\eqref{eq:kernel_size}, filtering with Gaussian kernel provides similar
distributions of filtered fields. 

We confine this analysis to the commutation error of the divergence and stress
tensor terms, which together give the \(\vec{F}^{\ell_V}\) source term. We
omit the commutation errors related to the derivatives present inside the
stress tensor, as is difficult to ensure that they are computed accurately,
due to non-constant viscosity in the LES equations. We also do not consider the
terms in the continuity equation, as the flow is assumed to be statistically
one-dimensional, which in turn necessitates a zero source term in the
continuity equation.

While performing a DNS combined with VANS equations, \citet{breugem2005}
assumed that the filter changes in the region where porosity increases from
\num{0.875} to \num{1}, that is for \(0.075H < y <0\). To investigate the
magnitude of those errors we prescribe the distribution of \(\ell_V\) using
Equation \eqref{eq:porosity_distr}, assuming that \((\delta^-,\delta^+) =
(0.075H, 0)\). The filter size changes from \(H/10\) inside the porous region to the value
of \(H/40 \approx 9.8^{+t}\), which is comparable to the resolution in the
centre of the channel, offered by the fine mesh used for particle-resolved LES.
The resultant \(\ell_V\) curve is visible in Figure~\ref{fig:bal_comm_err}. 

The momentum budget in the streamwise direction
including the commutation error can be given as 
\begin{equation}
  \begin{split}
  0 =
  \underbrace{-\phi\dpd{\DAvg{p}}{x}}_\text{\textbf{I}}
  + \underbrace{\dpd{}{y}\sDAvg*{\nu_\mathrm{eff}\left(\dpd{u}{y} + \dpd{v}{x}\right)}}_\text{\textbf{II}}
  + \underbrace{\dpd{\phi(\DAvg{u}\DAvg{v} - \DAvg{uv})}{y}}_\text{\textbf{III}}
  + \underbrace{F_x}_\text{\textbf{IV}}\\
  + \underbrace{\dpd{\ell_V}{y}\left\{
      \dpd{G}{\ell_V} \star \left( 
      \gamma uv - \gamma\nu_\mathrm{eff}\left(\dpd{u}{y} + \dpd{v}{x}\right)\right)
  \right\}}_\text{\textbf{V}}, 
  \end{split}
  \label{eq:mom_balance}
\end{equation}
with the terms \textbf{I}-\textbf{V} denoting effective pressure gradient, the contribution
of viscous and sub-filter stresses, drag force and commutation error respectively.
The commutation error related to the pressure gradient is equal to zero, as 
the gradient of \(\ell_V\) is aligned with the wall-normal direction. 

The accuracy of the computation of the commutation error is tested by
evaluating Equation \eqref{eq:mom_balance}. The distribution of the terms is shown
in Figure~\ref{fig:bal_comm_err} along with the curves computed with the
constant filter size. Non-uniform filtering width introduces an error by
changing the contributions of different terms to the momentum balance.
Decreasing filter size leads to a faster change of the drag force and an abrupt
change of the derivative of the viscous and sub-filter stresses. 
This behaviour adds validity to the discontinuous method of modelling the
interface, where the stress jump is prescribed at the coupling porous-fluid
boundary (see e.g.~\cite{silva2003} for this method applied to the channel
flow). In such an approach the filtering size is different on both sides of the
boundary. Assuming that \(\ell_V\) changes continuously over such an interface,
albeit with a steep gradient, it has to result in a commutation error and an
abrupt change in the stresses. Therefore, imposing a stress jump at this
location can be seen as a method to account for some part of the commutation
error.

When included in the momentum budget, the commutation error acts as a sink 
effectively increasing drag in the region where filtering width changes. 
Locally, its contribution is significant as it is comparable to drag.
However, it only constitutes \(2.5\%\) of the total combined source term 
in the momentum equation, in the \(-0.8\geq y/H\geq 0.8\) region.

\begin{figure}[t]
  \centering
  \subfloat[
      commutation error related to the convection term, given by 
      \((\partial \ell_V /\partial x_i) (\partial G / \partial \ell_V )\star (\gamma u_i u_j)\)
  ]{
    \includegraphics[width=0.47\textwidth]{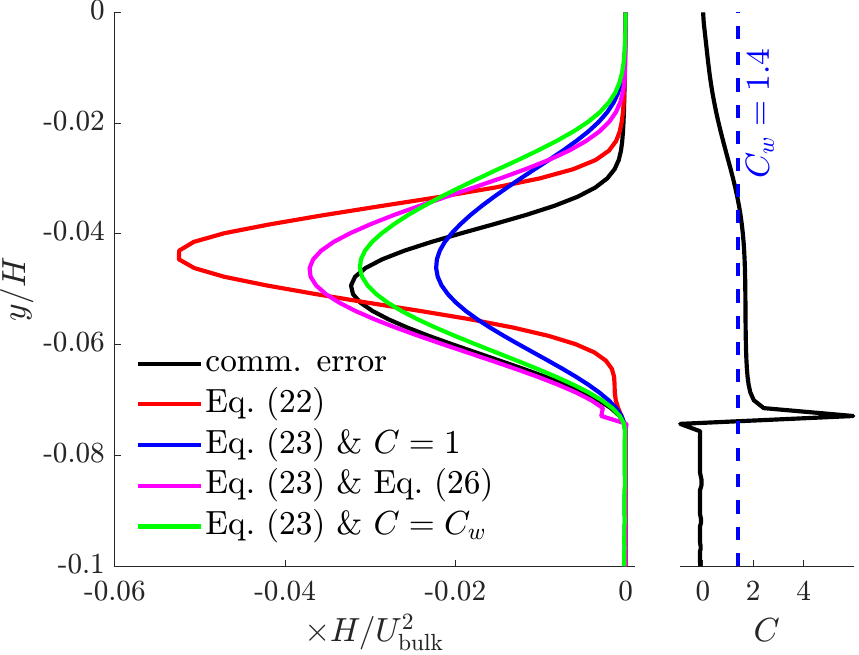}
    \label{fig:uu_comm_err}
  }
  \hfill
  \subfloat[
    commutation error related to the convection term, given by 
    \((\partial \ell_V /\partial x_i) (\partial G / \partial \ell_V )
    \star (-\gamma \nu_\mathrm{eff}( \partial u_i/\partial x_j + \partial u_j/\partial x_i))\)
  ]{
    \includegraphics[width=0.47\textwidth]{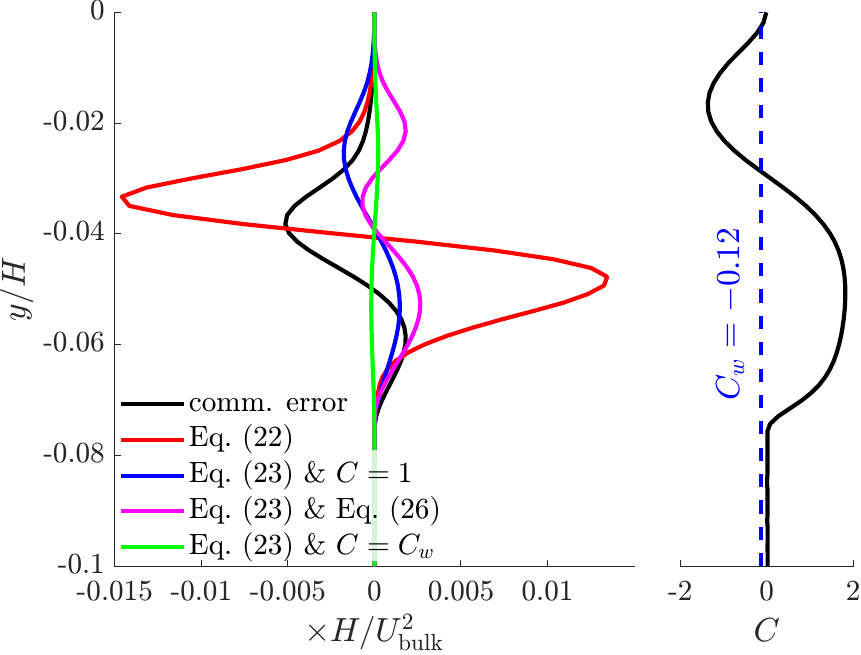}
    \label{fig:vis_comm_err}
  }
  \caption{
    Commutation errors assuming distribution of filter size drawn in
    Figure~\ref{fig:bal_comm_err}. It is compared to an approximation of the CE
    given by Equations~\eqref{eq:error_approx} and the scale-similarity model.
    Three different methods for specifying the model coefficient are presented.
    The distribution of the coefficient \(C\) computed from
    Eq.~\eqref{eq:C_ss_coeff} along with the value of its average weighed by
    \(|\partial \ell_V /\partial x_i|\) (blue dash line) are visible on the
    right.
  }
\end{figure}

The commutation error arising from filtering \(\partial u_i u_j /\partial x_j\)
constitutes the main contribution to \(F^{\ell_V}_x\) (Figure
\ref{fig:uu_comm_err}). The value obtained from the particle-resolved simulation has
been compared to the models presented in Section~\ref{s:models}. Equation
\eqref{eq:error_approx} does not provide a good prediction of the error
distribution. On the other hand, an approach based on the scale-similarity
hypothesis provides a much better estimation of the commutation error. If the
coefficient in Equation \eqref{eq:ss_model} is chosen as \(C=1\), the model is
relatively inaccurate, however, if the distribution of \(C\) is determined
according to Equation \eqref{eq:C_ss_coeff}, it performs much better,
especially for the lower portion of the region in which filter changes size.
That being said, the value of \(C\) changes drastically around \(y=-0.075H\). 
In the world of dynamic LES models, such behaviour is usually remedied by
averaging the coefficient in homogeneous directions~\cite{sagaut2006}. Following
this method, \(C\) was determined by averaging the values obtained
from Eq. \eqref{eq:C_ss_coeff} with the absolute value of \(\partial \ell_V /
\partial y \) as a weight:
\begin{equation}
  C_w = \int_{-H}^{H}{C\left|\dpd{\ell_V}{y}\right|}{\dif\,y}
  {\Bigg(\int_{-H}^{H}{\left|\dpd{\ell_V}{y}\right|}{\dif\,y}\Bigg)}^{-1}.
  \label{eq:weighted_c}
\end{equation}
This resulted in a value of \(C_w = 1.4\). 
the model with \(C=C_w\) has the best overall performance, reproducing accurately the values
of the error for the lower portion of the interface while limiting the overprediction 
for higher \(y\) coordinates.

The error related to the divergence of the stress tensor (Figure
\ref{fig:vis_comm_err}) constitutes a much smaller part of the overall commutation
error. This is an expected result, as the sub-filter stresses and drag force
have the most substantial contributions to the momentum budget. Each of the tested
models was not able to predict the distribution of the error with sufficient
accuracy. The bad performance of the
scale-similarity model stems from the computed distribution of \(C\), which
changes the sign in the middle of the interface region. This also results in a near zero
value of the weighted average \(C_w\). The bad performance of both methods for
this particular term is discouraging, however, since the commutation error
related to the viscous term has little impact on the momentum balance, it can
be safely neglected while the modelling of CEs is attempted, without impeding the
accuracy of results.

\subsection{Solving DANS equations: case G2}

The solutions obtained on the coarse meshes did not deviate
much from the ones computed on the finer ones. Each of the simulations 
resolved more than \SI{95}{\percent} of turbulent kinetic energy locally,
therefore the resolution of the coarser mesh can be deemed sufficient for such
simulations. In the following section, the results from the finer meshes will be
presented.

\begin{figure}[t]
  \begin{center}
    \includegraphics[width=0.95\textwidth]{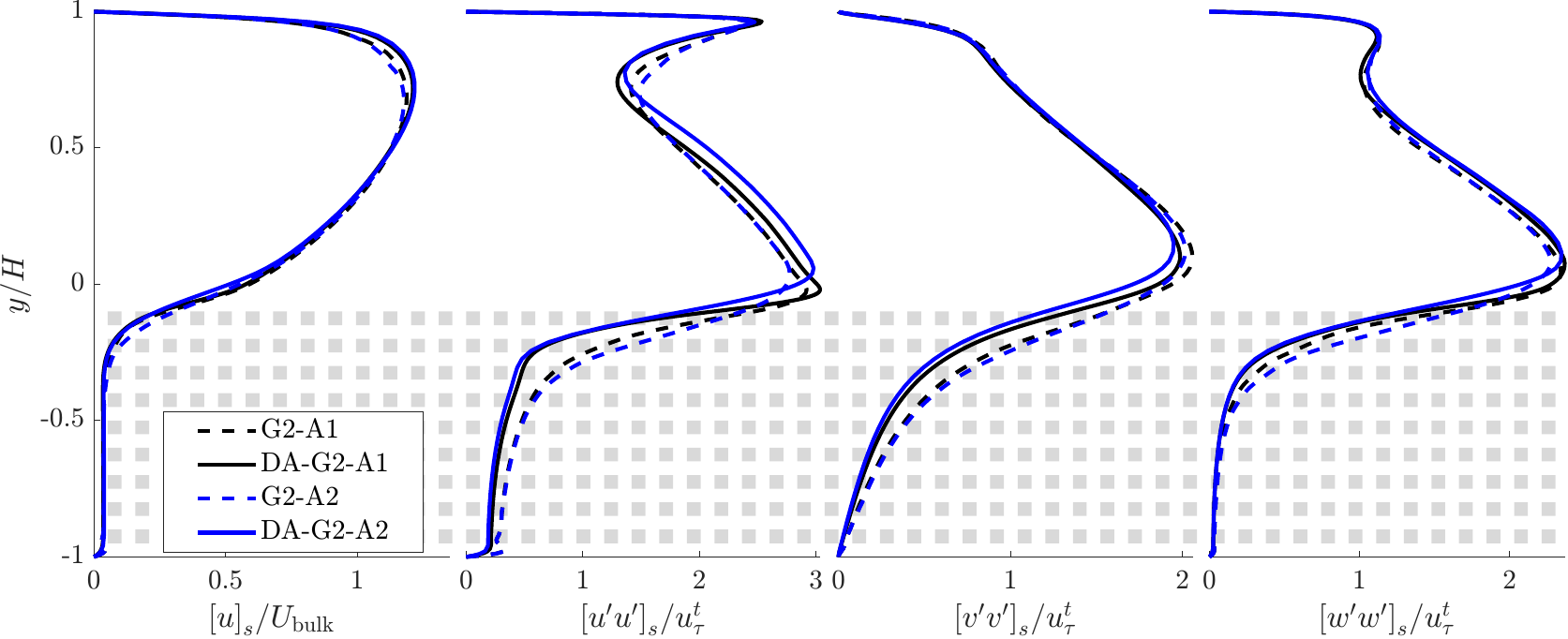}
    \caption{Comparison of the profiles from the double-averaged
    particle-resolved LES and LES with DANS equations results averaged in time. The
  results are presented for two different filter sizes, for G2-A1 case \(\ell_V=H/10\) and 
  for G2-A2 \(\ell_V=H/5\).}
  \label{fig:channel_means_va}
  \end{center}
\end{figure}

Considering that an \emph{a-posteriori} analysis of the commutation error
performed in Section~\ref{s:comm_G2} indicated that the CE does not have a
detrimental effect on the momentum balance, it was neglected in DA-G2
simulations. As in the case of \citet{breugem2005}, who performed a DNS, a good
correspondence between the solution of modelled equations and the results of PRS
can be observed. Focusing first on results of DA-G2-A1 (assuming \(\ell_V =
H/10\) the profile of mean velocity in the clear section of the channel (Figure
\ref{fig:channel_means_va}) is slightly shifted towards the upper wall. The
mean velocity in the porous and interface regions is also predicted with
sufficient accuracy. Figure~\ref{fig:channel_means_va} also presents resolved
normal fluctuations computed in all directions. Due to the fact that time and
spatial filtering can be applied in an interchangeable order, the
double-averaged fluctuation can be computed as 
\({\langle u_i^\prime\rangle} = \overline{\langle u_i\rangle} - {\langle u_i\rangle}\).
This results in the following definition of the components of double-averaged
resolved fluctuations:
\begin{equation}
  \sDAvg{u^\prime_i u^\prime_j} 
  = \phi\overline{\langle u_i^\prime\rangle  \langle u_j^\prime\rangle}
  + \phi\left( 
       \overline{\langle u_i^\prime u_j^\prime \rangle} 
    -  \overline{\langle u_i^\prime\rangle  \langle u_j^\prime\rangle}
  \right)
  \label{eq:fluctuations}
\end{equation}
The term enclosed in parenthesis on the right-hand-side cannot be 
directly computed from the solution of filtered equations, therefore, 
it has to be neglected. 

The components of the resolved TKE are consistently underestimated in the porous
part of the domain. On the other hand, they are accurately predicted near the
top wall, in the boundary layer. The fluctuations near the
topmost layer of cubes are also computed with reasonable accuracy.
All of the above suggests, that the chosen drag closure is a viable method to
be used with an LES model. Obtained solutions offer similar fidelity as the
reference DNS \cite{breugem2005}. 

The increase of assumed filter size leads to the change of the explicitly
filtered velocity profile at the interface, with the higher velocities
penetrating the porous structure deeper. This change is not reproduced in the
results obtained with the implemented solver. The velocity profile is only
slightly affected by the change in filter size. However, it does have an effect
on the second-order statistics, as the increase of \(\ell_V\) shifts the peak
of resolved fluctuations in the lower boundary layer upwards.

\begin{figure}[t]
  \begin{center}
    \includegraphics[width=0.5\textwidth]{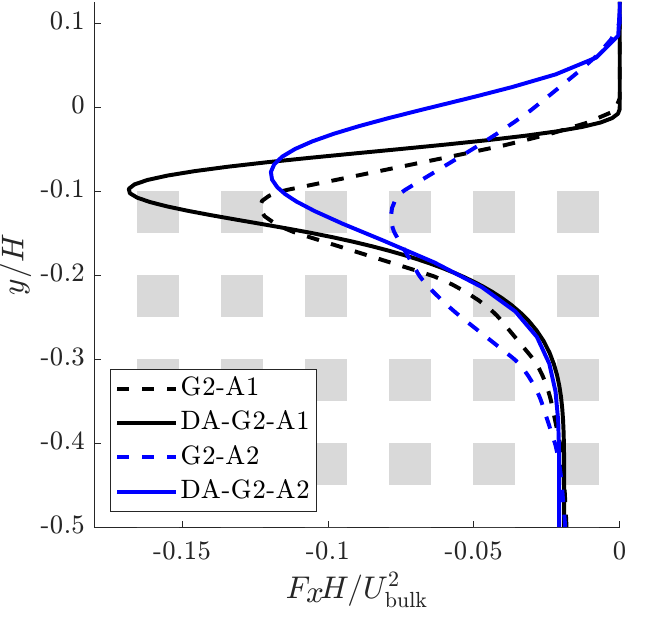}
    \caption{Comparison of the drag force from the filtered PRSs and LES done
    with the DANS solver. The results are presented for two different filter
    sizes, \(\ell_V=H/10\) and \(H/5\).}
  \label{fig:channel_drag_va}
  \end{center}
\end{figure}

The profiles of drag force around the interface region are presented 
in Figure~\ref{fig:channel_drag_va}. For both filter sizes, the magnitude 
of the force is overestimated. In fact, comparing total drag force from 
explicitly filtered PRSs and the drag model
\(
  F_x =
  -\nu \mathcal{K}^{-1}\left(1 + \mathcal{F}\right)\phi^2\DAvg{u},
\)
the simulation assuming \(\ell_V=H/10\) predicts \(1.1\) greater drag. The
accuracy of the model seems to decrease with increasing filter size, as for
\(\ell_V=H/5\), this ratio is equal to \(1.21\). In the model, the force is
proportional to the predicted velocity. For \(\ell_V=H/10\) the model resembles
the shape of the drag curve obtained from the PRS much more closely, which
results in a slightly more accurate velocity prediction. In both cases, the
magnitude of the drag force curve rises in a similar fashion while approaching the 
interface from below, which explains why both filter sizes ultimately result in
similar velocity distributions in this region.

The change in the porosity at the interface has a limited effect on the overall
performance of the model. However, if the flow near the porous-fluid interface
is to be investigated, then the profile of \(\phi\) has to be selected
carefully. For example, if Darcy--Forcheimer closure is chosen then \(\phi\)
might be determined in such a way that generates the best approximation of
\(\vec{F}\) at the interface. This can naturally only be achieved if the
results of a more accurate, particle-resolved simulation are available.
Ultimately, as also pointed out by \citet{sadowski2023}, the fact that the
change of \(ell_V\) has an influence on the accuracy of drag force computation
indicates that the closure should include the influence of that parameter, either by 
correcting \(\mathcal{K}\) and \(\mathcal{F}\) or other means.

\section{Concluding remarks}\label{s:conclusions}

In the present study, the mathematical foundations of the double-averaging framework
for flows in porous media have been rigorously reformulated as filtering with a
space-time kernel. The averaging theorems used extensively in the derivation of
Double-Averaged Navier Stokes equations have been generalised to work with any
well-behaved kernel functions and extended to include commutation errors of
filtering/derivative operators.

This development lays the groundwork for using DANS equations with inhomogeneous
filtering, an important property of the equations for flow configurations where
significant changes in the pore or particle size occur. Additionally, relationships
describing the errors arising when filtering next to a boundary are derived and
included in the equations. These errors have fundamentally the same form as
boundary integrals forming the unclosed drag force terms.

The newly derived equations are tested by explicitly filtering a simplified
particle-resolved simulation with a Gaussian filter and evaluating the residual
of both momentum and mass conservation. Although only spatial filtering has
been conducted, this has proved sufficient to verify the derivation of
mathematical description of the errors. The largest relative momentum error
(around \(1\%\)) has been found in the inhomogeneous filtering case, suggesting
that extensive analysis of grid requirements for a given distribution of filter
width might be necessary.

Furthermore, the particle-resolved LES of the fully developed turbulent flow 
in a channel partly occupied by a porous structure has been realised. The 
computation has been performed on two meshes, testing the grid requirements 
for an accurate scale-resolving simulation in such a setting. Insufficient 
resolution near the surface of the porous wall led to inaccurately predicted 
mean velocity in the whole porous region of the channel and greatly overestimated 
porous-induced drag force, even though, the results in the unobstructed region 
agree qualitatively with the reference DNS. 

The results from the LES have also been explicitly filtered. The assumed
distribution of the kernel molecule was non-uniform, transitioning from a size
adequate to homogenise the porous medium to a smaller filter in the
unobstructed region, representing the filtering requirements for an LES. Resulting
commutation error have been computed and compared to other terms contributing
to the momentum balance. In such a configuration, the error acts as an additional sink 
term in the momentum equation, effectively increasing the drag force, albeit by 
a small margin, as it only constituted around 2.5\% of the total effective drag.

The bulk of the commutation error is related to the derivative of the
convective term in the Navier-Stokes equations. The error manifests itself as a
decrease in the magnitude of the drag force and a local abrupt change in the
divergence of the sub-filter and viscous stresses. This behaviour adds validity
to the discontinuous treatment of the porous-fluid interface with the stress
jump boundary conditions. Such coupling can be viewed as a method of modelling
the commutation error arising from a discontinuous change in the filter size.

Two different approaches for modelling the commutation error have been
investigated, based on the differential approximation of the filtering
operation and the scale-similarity hypothesis. The first one, while easier to
implement did not yield good results. The latter, when the model constant has
been averaged, was able to qualitatively reproduce the error related to the
convective term.

A solver for double-filtered equations, based on the
PIMPLE algorithm has been implemented in OpenFOAM. The numerical scheme can
reproduce the explicitly filtered results with very good accuracy, employing
source terms gathered from explicit filtering. Additionally, the
significance of the error source terms has been investigated for the accurate
computation of filtered velocity and pressure. This analysis revealed that the
inclusion of boundary commutation errors is not necessary in both the momentum
and continuity equation when the porosity distribution is properly prescribed in
those areas. 

Finally, the channel flow has also been simulated with a combined LES porous-medium model,
which assumed eddy-viscosity closure in the whole domain. The porosity was
prescribed to resemble the profile obtained from the explicitly filtered
results, considering two different filter sizes. Following the conclusions,
that commutation errors might not have a critical impact on the results in such
configuration, they have been neglected. The solution was found to be in good
agreement with explicitly filtered particle-resolved LES results. However, computed fields
did not reflect the rather subtle changes of filtered quantities at the
interface, occurring with an increase in filter size. This seems to be a result
of the inaccuracy of the drag closure which overpredicted the total drag by 
10\% for the smaller of the chosen filter sizes and 20\% for the larger one.

%

The presented mathematical framework opens doors to several possibilities for
future research. First and foremost a full-scale flow configuration e.g.\
gaseous flow in a lime kiln could be evaluated using explicit filtering.
The forces, dispersion stresses, turbulent quantities and filtered scalar
fluxes could be compared to the currently employed models. Even though the
investigated numerical method resulted in a qualitatively accurate prediction of
the flow at the fluid-porous interface, it is important to stress that the drag
closure was tuned to the present geometrical configuration (i.e.\ array of
cubes). Therefore, gathering more data from particle-resolved simulations
should aid in the derivation of a more general approach for drag modelling.
Additionally, the investigation of the drag closure seems like a more promising
avenue to increase the fidelity of the simulation of porous-fluid systems, as it
contributes the biggest error to the momentum balance.


\section*{Conflict of Interest}
The authors have no conflicts to disclose.

\section*{Author's Contributions} \textbf{Wojciech Sadowski:} conceptualization
(equal), data curation, formal analysis (lead), writing/original draft
preparation (lead), methodology, software, validation (lead), visualization
(lead), writing/review \& editing (equal). \textbf{Mohammed Sayyari:} formal
analysis (supporting), validation (supporting), visualisation (supporting),
writing/original draft preparation (supporting), writing/review \& editing
(equal). \textbf{Francesca di Mare:} conceptualization (equal), formal analysis
(supporting), funding acquisition, project administration, supervision (lead),
writing/review \& editing (equal). \textbf{Holger Marschall:} conceptualization
(equal), writing/review \& editing (equal), writing/original draft preparation
(supporting).

\section*{Acknowledgement}
The authors gratefully acknowledge financial support from the Deutsche
Forschungsgemeinschaft (DFG) thorough SFB/TRR287, Project  Number 422037413.
The authors also wish to thank Dr.\ Pascal Post for the fruitful discussions and
suggestions given during preparation of this manuscript.

\appendix

\section{The superficial space-time averaging technique}
\label{s:classical_averaging}

The superficial space-time average \citep{nikora2013} of variable
\(\psi(\vec{x}, t)\), denoted as \(\sDAvg{\psi}\), is defined as
\begin{equation}
  \sDAvg{\psi}(\vec{x}, t) = 
  \dfrac{1}{T_0 V_0}\regionint{T_0 \times V_0}
  {\psi(\vec{x} + \vec{\xi}, t +\tau)\gamma(\vec{x} + \vec{\xi}, t +\tau)}
  {\dif\vec{\xi}\dif\tau}.
\end{equation}
The function \(\gamma\) is often called a clipping or phase indicator function
\citep{nikora2013, breugem2006}. It describes the distribution of
the fluid phase in space and time,
\begin{equation}
  \gamma(\vec{x}, t) =%
  \begin{cases}
    1, &  \text{if the particle at \(\vec{x},t\) is fluid;}\\
    0, &  \text{otherwise.}
  \end{cases}
\end{equation}

The superficial average and the space-time porosity \(\phi_{VT}\) are related by the
formula for the intrinsic average denoted with \(\DAvg{\cdot}\) 
\begin{equation}\label{eq::sup_int_relation_appendix}
  \sDAvg{\psi} = \phi_{VT}\DAvg{\psi},
\end{equation}
where the space-time porosity is defined as
\begin{equation}
  \phi_{VT} = \sDAvg{1} =
  \dfrac{1}{T_0 V_0}\regionint{T_0 \times V_0}
  {\gamma(\vec{x} + \vec{\xi}, t +\tau)}
  {\dif\vec{\xi}\dif\tau}.
\end{equation}

To obtain a space-time averaged system of differential equations, the
superficial averages are applied to the time and space partial derivatives
respectively. Next the averaging theorems are used to introduce the averaging
operators under the derivatives. Initial derivations of these theorems for VANS
with the discussion of their properties can be found in works by
\citet{whitaker1985} and \citet{howes1985}. \citet{gray1977} followed a
different approach to prove both theorems, using the phase indicator function
\(\gamma\) and its derivatives \citep{kinnmark1984}, which was later extended
by \citet{nikora2007a, nikora2013} to obtain the relations presented above. The
main assumption in each proof is that \emph{the averaging volume remains
space-invariant}. The theorems describe the relationship between the average of
the derivative and the derivative of the average in the following way:
\begin{equation}\label{eq::da_theorem_t}
  \sDAvg*{\dpd{\psi}{t}}
  = \dpd{\sDAvg{\psi}}{t} + \frac{1}{V_0} \overline{%
    \oint\limits_{S} \psi w_i n_i
  \;\dif S }^s,
\end{equation}
\begin{equation}\label{eq::da_theorem_x}
  \sDAvg*{\dpd{\psi}{x_i}} = 
  \dpd{\sDAvg{\psi}}{x_i}  - \frac{1}{V_0} \overline{%
    \oint\limits_{S} \psi n_i
    \;\text{d} S }^s.
\end{equation}
\(S\) is the interface
region between the fluid and solid phases inside averaging volume \(V_0\),
\(\vec{n}\) is the inward normal vector of the fluid region and \(\vec{w}\) is
the velocity of the interface. Overbar with subscript \(s\) denotes superficial
time average.
Using equation \eqref{eq::sup_int_relation}, both theorems can be used with
intrinsic quantities.

\section{Derivation of DANS equations}
\label{s:dans_app}

We start by averaging the system~\eqref{eq:ins} in space and time following
\citet{nikora2013}. Using Equations~\eqref{eq::da_theorem_t}
and~\eqref{eq::da_theorem_x} results in the following momentum equation:
\begin{equation}\label{eq::sda_ns_mom}
  \rho\left(\dpd{\sDAvg{u_i} }{t} + \dpd{ \sDAvg{ u_i u_j} }{x_j}\right) = \dpd{\sDAvg{\sigma_{ij}}}{x_j} +
  \rho\sDAvg{f_i} + \rho F_i \;,
\end{equation}
where \(\vec{F}\) represents the interfacial forces forces introduced with
surface integrals while applying averaging theorems:
\begin{equation}
  \label{eq::porous_drag}
  F_i = -\dfrac{1}{V_0}\overline{\oint\limits_{S} \sigma_{ij} n_j \dif S }^s .
\end{equation}

Equation \eqref{eq::sda_ns_mom} is an intermediate step for obtaining DANS momentum equation.  
The next step requires changing the averages into their intrinsic counterparts.
The choice of writing the equation with intrinsic averaging has two advantages. 
First, as shown by \citep{whitaker1999}, intrinsic averaging preserves constants, 
where superficial averaging does not. 
Second, the intrinsic average results in a proper decomposition of the convection
term. The dispersion terms resulting from performing such decomposition assuming 
superficial average cannot be modelled by a diffusion-like mechanism \citep{gray1975}.
In both cases, an improper choice of the averaging type may result in an error
that is of the order of magnitude of the porosity. Additionally, inconsistent formulation 
of the DANS system with superficial variables may lead to non-Galilean invariant set 
of equations~\citep{wang2015}. We will employ intrinsic averages in the course of this work, 
however, it is important to note that, if used with care, superficial variables can be 
used with success and accuracy \citep{breugem2005, silva2003}. 

Rewriting Equation~\eqref{eq::sda_ns_mom} using the intrinsic average results in
\begin{equation}\label{eq::sda_ns_mom_i}
  \rho\left(\dpd{\phi_{VT}\DAvg{u_i} }{t} + \dpd{ \phi_{VT}\DAvg{ u_i u_j} }{x_j} \right)= 
  \dpd{\phi_{VT} \DAvg{\sigma_{ij} }}{x_j} + \rho\phi_{VT}\DAvg{f_i} + \rho F_i.
\end{equation}
In the second term in the left-hand side of Equation \eqref{eq::sda_ns_mom_i}, the 
average of the product of velocities must be decomposed:
\begin{equation}\label{eq::atau}
  \DAvg{u_i u_j}  = \DAvg{u_i}\DAvg{u_j} - \tau_{ij} 
                  = \DAvg{u_i}\DAvg{u_j} - (\DAvg{u_i}\DAvg{u_j} - \DAvg{u_i u_j}),
\end{equation}
where \(\tnsr{\tau}\) represents porous dispersion and must be modeled, if its
influence on the behaviour of the flow is to be included in the mathematical
derivation. However, it is often the case that this term is neglected in the final
equations (analysis of the significance of \(\tnsr{\tau}\) in the context of
VANS framework can be found in works of \citet{breugem2006}). The averaged 
continuity equation is derived in a similar fashion.
The result is that the space-time averaged Navier-Stokes equations:
\begin{subequations}
  \begin{eqnarray}
  \dpd{\phi_{VT}}{t} + \dpd{ \phi_{VT} \DAvg{u_i} }{x_i} &=& 0,\\
	  \rho\left(
      \dpd{\phi_{VT}\DAvg{u_i}}{t} + \dpd{\phi_{VT}\DAvg{u_i}\DAvg{u_j}}{x_j}
  \right)  &=&
    \dpd{\phi_{VT} \DAvg{\sigma_{ij} }}{x_j}
       + \dpd{\rho\phi_{VT} \tau_{ij} }{x_j}
     + \rho\phi_{VT}\DAvg{f_i} + \rho F_i,
  \end{eqnarray}
\end{subequations}

\section{Derivatives for evaluation of commutation errors}\label{s:app_drel}
This section outlines formulas for derivatives, necessary for simplification of equation for the 
commutation error of space-time filter.
First, the derivative of a filter kernel is given by (\(s\) is a placeholder  for \( x_i\) or \(t\))
\begin{eqnarray}
  \dpd{G(\vec{\xi}, \tau, \ell_V(\vec{x}), T_0(t))}{s} 
  = \dpd{G}{\xi_i}\dpd{\xi_i}{s}
  + \dpd{G}{\tau}\dpd{\tau}{s}
  + \dpd{G}{\ell_V}\dpd{\ell_V}{x_i}\dpd{x_i}{s}
  + \dpd{G}{T_0}\dpd{T_0}{t}\dpd{t}{s}\;,
\end{eqnarray}
which leads to following simplifications under the assumptions made in section \ref{s:comm_err}:
\begin{subequations}\label{eq::G_derivative}
  \begin{eqnarray}
  \dpd{G(\vec{\xi}, \tau, \ell_V(\vec{x}), T_0(t))}{x_i} 
  = \dpd{G}{\ell_V}\dpd{\ell_V}{x_i},\\
  \dpd{G(\vec{\xi}, \tau, \ell_V(\vec{x}), T_0(t))}{t} 
  = \dpd{G}{T_0}\dpd{T_0}{t}
  \end{eqnarray}
\end{subequations}

The spatial derivative of the clipping function \citep{gray1977, howes1985} can be defined as
\begin{equation}\label{eq::gamma_derivative_1}
  \dpd{\gamma}{x_i} = n_i \delta(\vec{x} - \vec{x}_S)\;,
\end{equation}
where \(\delta\) is the Dirac function, \(\vec{x}_S\) is a vector ``tracing'' the
interface \(S\) and \(\vec{n}\) is a normal vector defined as in section \ref{s:comm_err}.
The time derivative of \(\gamma\) can be found after noticing that the material derivative of \(\gamma\) is
zero for an observer sitting on the iso-surface (respectively iso-line) of 
\(\gamma = 1\) (i.e.\ where \(\vec{x} = \vec{x}_S\)). Owing to that 
\begin{equation}\label{eq::gamma_derivative_2}
  \dpd{\gamma}{t} = - w_i\dpd{\gamma}{x_i} = w_i n_i \delta(\vec{x} - \vec{x}_S),
\end{equation}
where \(\vec{w}\) is the velocity of the interface.
For completeness, the derivatives of the shifted functions under the convolution integral can be
obtained in the following manner:
\begin{equation}
  \dpd{f(\vec{x} - \vec{\xi}, t - \tau)}{s} =
  \dpd{f}{(x_i - \xi_i)}\dpd{(x_i - \xi_i)}{s}
  + \dpd{f}{(t - \tau)}\dpd{(t - \tau)}{s},
\end{equation}
leading to:
\begin{subequations}\label{eq::f_derivative}
  \begin{eqnarray}
    \dpd{f(\vec{x} - \vec{\xi}, t - \tau)}{x_i} &=& \dpd{f}{(x_i - \xi_i)},\\
    \dpd{f(\vec{x} - \vec{\xi}, t - \tau)}{t} &=& \dpd{f}{(t-\tau)}
  \end{eqnarray}
\end{subequations}

\section{Sensitivity study of explicit filtering w.r.t. kernel clip distance
\(d\) and filtering mesh resolution}\label{s:sens_study}
Explicit filtering method, as described in \cite{sadowski2023}, can have multiple 
sources of errors, that have to be considered and controlled while analysing 
filtered quantities. Firstly, if the mesh is to coarse to accurately represent
the chosen filter, this can introduce an error due to insufficient
resolution of numerical integration. Secondly, if the spatial resolution differs
greatly between the two parts of the domain, the difference in accuracy of convolution computation
will introduce oscillations in the filtered fields.

Moreover, as described in the section \ref{s:numerics} the filter kernel is being clipped
to 0 after a certain distance \(d\) from the center of the filtering molecule.
This can introduce additional significant error to the filtering operation. 
A clipped kernel, which originally had an infinite support, does
not conserve constants, which also means that the filtered flow will be
wrongly represented. 

All of those inaccuracies will accumulate and influence the momentum balance.
For example, oscillations in porosity field resulting from the errors, will be
carried over to other quantities of interest and increased by the use discrete
derivative operators, e.g.\ when evaluating \(\text{div}\;( \phi \DAvg{\vec{u}}
\DAvg{\vec{u}})\).

\begin{figure*}
  \centering
  \subfloat[Gaussian filter]{
    \includegraphics[width=0.45\linewidth]{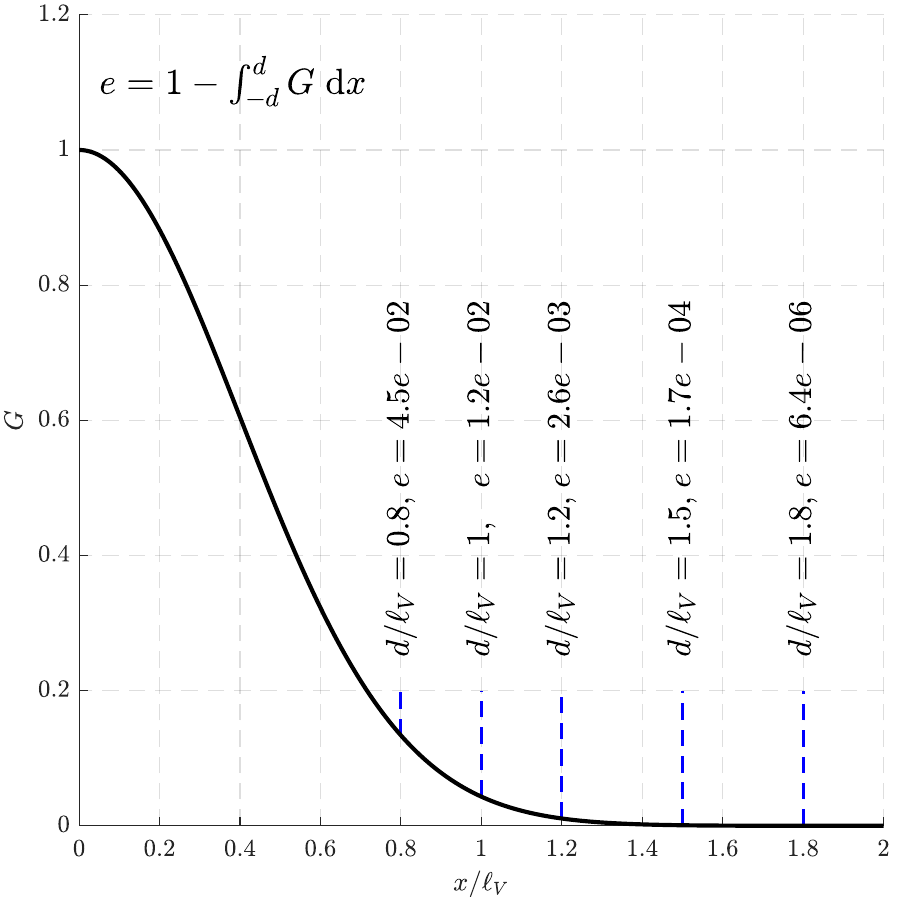}
    \label{fig:Gaussian_sensitivity}
  }
  \hfill
  \subfloat[derivative of the Gaussian filter with respect to filter width \(\ell_V\)]{
    \includegraphics[width=0.45\linewidth]{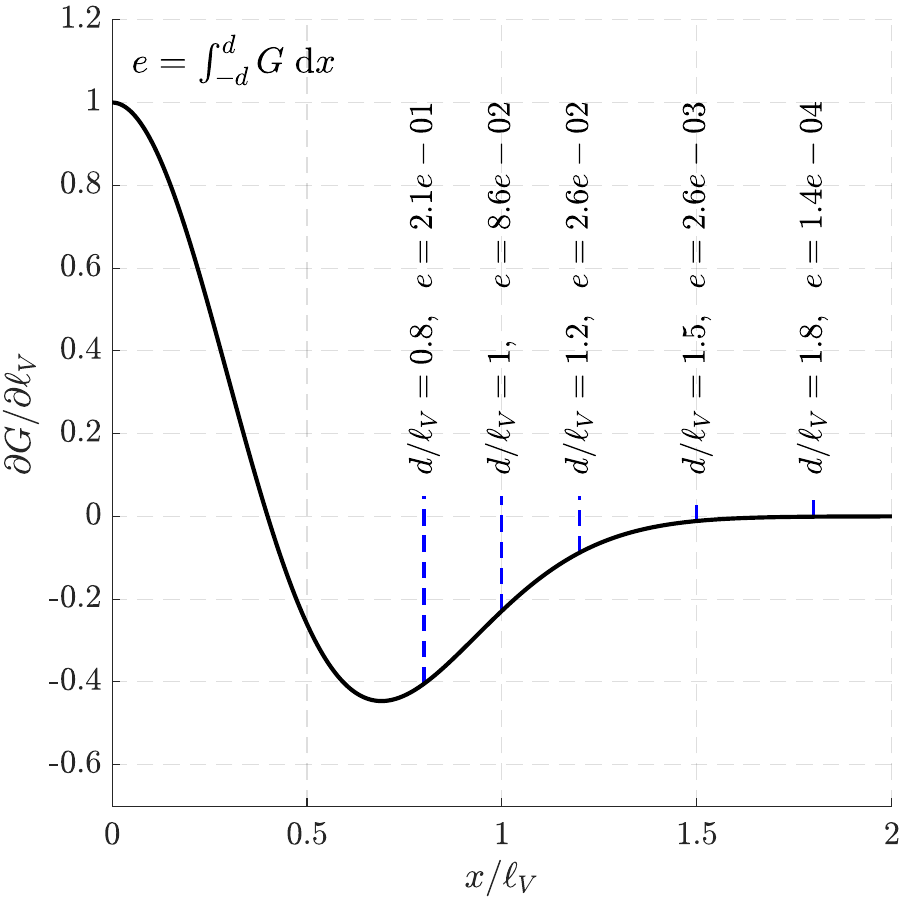}
    \label{fig:GaussianDer_sensitivity}
  }
  \caption{Depiction of the clipped filter shapes used in the sensitivity
    study. At chosen distance from the center \(d/\ell_V\) kernel is clipped to
    0. This threshold is shown in the figures by the blue dashed lines. Each
    value of  \(d/\ell_V\) is paired with corresponding value of the clipping
    error. 
  }
\end{figure*}

To test what value of \(d\) is required for accurate representation of the
filter, a sensitivity study was performed with a set of distances \(d/\ell_V =
\{0.8, 1, 1.2, 1.5, 1.8\}\) using both cases G1-A and G1-C. The corresponding
filter shapes (the Gaussian filter and its derivative with respect to filter
width \(\ell_V\)) are drawn on Figures \ref{fig:Gaussian_sensitivity} and
\ref{fig:GaussianDer_sensitivity} in their one-dimensional forms. The shapes of
the both kernels alone, suggest that latter will require greater value of
\(d/\ell_V\) to ensure proper resolution.

\begin{figure*}
  \centering
  \subfloat[case G1-A]{
    \includegraphics[width=0.47\linewidth]{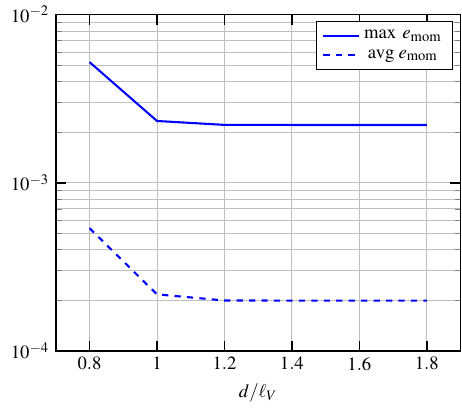}
    \label{fig:fil_sens_case_A}
  }
  \hfill
   \subfloat[case G1-C]{
    \includegraphics[width=0.47\linewidth]{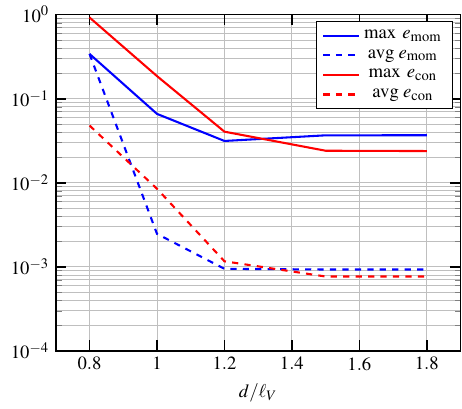}
     \label{fig:fil_sens_case_C}
   }

  \caption{Values of the maxima and averages of the residuals of momentum
    \(e_\text{mom}\) and continuity equation \(e_\text{con}\) of DANS system.
    The residuals were computed evaluating the balance of explicitly filtered 
    terms of the both equations and are plotted against the filter clipping
    distance \(d/\ell_V\).}
\end{figure*}

For each value of \(d/\ell_V\) the cases were filtered and residuals of 
double-filtered equations were evaluated as described in section
\ref{s:results}. Obtained error values are plotted in the figures
\ref{fig:fil_sens_case_A} and \ref{fig:fil_sens_case_C}. Inspection of the errors
from case G1-A reveals that increasing the clipping distance beyond \(d/\ell_V =
1.2\) does not decrease the evaluated residuals, therefore this value has been
selected as a minimal clipping distance for filtering with Gaussian filter.

Based on the results from case G1-C, the minimal clipping distance for the
derivative was chosen to be \(d/\ell_V = 1.5\). Since the values of the
computed residuals did not decrease to values similar to case A, filtering mesh
(the mesh used for evaluation of balance of filtered quantities) sensitivity
study was additionally performed for case G1-C.

\begin{figure*}
  \centering
  \subfloat[residual values against the filtering mesh resolution, where
  \(\Delta\) denotes the element size and \(a\) is the size of the square]{
    \includegraphics[width=0.47\linewidth]{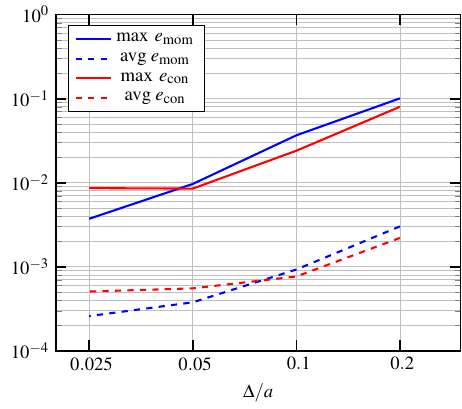}
    \label{fig:mesh_sens_case_C}
  }
  \hfill
  \subfloat[plots of the \(x\) component of the commutation error from
  inhomogenous filtering along the \(x\) axis in case G1-C]{
    \includegraphics[width=0.47\linewidth]{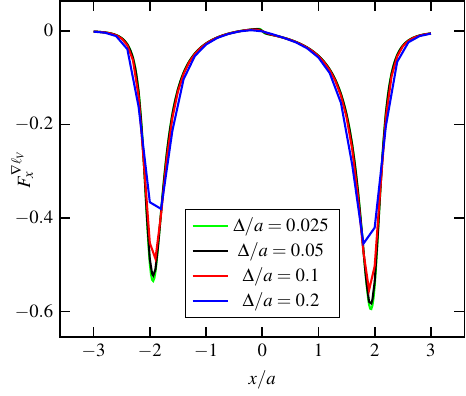}
     \label{fig:mesh_sens_case_C_plots}
   }

   \caption{Results of the filtering mesh sensitivity study.}
\end{figure*}

Results of the reference case were filtered on five uniform meshes with the
element size 
\[
  \frac{\Delta}{a}=\{0.2, 0.1, 0.05, 0.025\},
\]
where \(a\) denotes the length of the side of the square. The residuals were
again evaluated for every mesh and they are presented in Figure
\ref{fig:mesh_sens_case_C}. The mesh size of \(\Delta / a = 0.05\) was
selected as appropriate resolution for case G1-C as it reduced all error terms
below \(1\%\) and the evaluated source terms did not change significantly on
the finer mesh (Figure \ref{fig:mesh_sens_case_C_plots}). Interestingly the
maximum of continuity residual increased slightly when the finest mesh was
used. The reason for such behaviour is probably the fact that the resolution of
the filtering mesh is finer in most of the domain than the source mesh.


\pgfplotsset{
  log x ticks with fixed point/.style={
      xticklabel={
        \pgfkeys{/pgf/fpu=true}
        \pgfmathparse{exp(\tick)}%
        \pgfmathprintnumber[fixed relative, precision=3]{\pgfmathresult}
        \pgfkeys{/pgf/fpu=false}
      }
  },
  log y ticks with fixed point/.style={
      yticklabel={
        \pgfkeys{/pgf/fpu=true}
        \pgfmathparse{exp(\tick)}%
        \pgfmathprintnumber[fixed relative, precision=3]{\pgfmathresult}
        \pgfkeys{/pgf/fpu=false}
      }
  }
}

\bibliography{ms.bib}

\end{document}